\documentclass[floats,twocolumn,aps,prb,longbibliography,superscriptaddress]{revtex4-2}

\usepackage{graphicx}
\usepackage{amsfonts}
\usepackage{amssymb}
\usepackage{dsfont}
\usepackage{amsbsy}
\usepackage{amsmath}\usepackage[colorlinks = true,
            linkcolor = magenta,
            urlcolor  = magenta,
            citecolor = magenta,
            anchorcolor = magenta]{hyperref}
\usepackage{hyperref}
\usepackage{xcolor}
\usepackage{subeqnarray}

\begin{document}

\title{Magnetic properties of the spiral spin liquid and surrounding phases in the square lattice XY model}

\author{Mat\'ias G. Gonzalez}
\affiliation{Helmholtz-Zentrum Berlin f\"ur Materialien und Energie, Hahn-Meitner-Platz 1, 14109 Berlin, Germany}
\affiliation{Dahlem Center for Complex Quantum Systems and Fachbereich Physik, Freie Universität Berlin, 14195 Berlin, Germany\looseness=-1}

\author{Anna Fancelli}
\affiliation{Helmholtz-Zentrum Berlin f\"ur Materialien und Energie, Hahn-Meitner-Platz 1, 14109 Berlin, Germany}
\affiliation{Dahlem Center for Complex Quantum Systems and Fachbereich Physik, Freie Universität Berlin, 14195 Berlin, Germany\looseness=-1}

\author{Han Yan}
\affiliation{Institute for Solid State Physics, The University of Tokyo, Kashiwa, Chiba 277-8581, Japan}
\affiliation{Department of Physics and Astronomy, Rice University, Houston, TX 77005, USA\looseness=-1}
\affiliation{Smalley-Curl Institute, Rice University, Houston, TX 77005, USA}

\author{Johannes Reuther}
\affiliation{Helmholtz-Zentrum Berlin f\"ur Materialien und Energie, Hahn-Meitner-Platz 1, 14109 Berlin, Germany}
\affiliation{Dahlem Center for Complex Quantum Systems and Fachbereich Physik, Freie Universität Berlin, 14195 Berlin, Germany\looseness=-1}

\begin{abstract} 
Spiral spin liquids possess a subextensively degenerate ground-state manifold, represented by a continuum of energy minima in reciprocal space. Since a small change of the spiral state wavevector requires a global change of the spin configuration in real space, it is {\it a priori} unclear how such systems can fluctuate within the degenerate ground state manifold. Only recently it was proposed that momentum vortices are responsible for the liquidity of the spiral phase and that these systems are closely related to an emergent rank-2 U(1) gauge theory [\hyperlink{https://doi.org/10.1103/PhysRevResearch.4.023175}{H. Yan and J. Reuther, Phys. Rev. Research 4, 023175 (2022)}]. As a consequence of this gauge structure, four-fold pinch-point singularities were found in a generalized spin correlator. In this article, we use classical Monte Carlo and molecular dynamics calculations to embed the previously studied spiral spin liquid into a broader phase diagram of the square lattice XY model. We find a multitude of unusual phases and phase transitions surrounding the spiral spin liquid such as an effective four-state Potts transition into a colinear double-striped phase resulting from the spontaneous breaking of two coupled $\mathds{Z}_2$ symmetries. Since this phase is stabilized by entropic effects selecting the momenta away from the spiral manifold, it undergoes a re-entrance phenomenon at low temperatures into a nematic spiral phase. We also observe a region of parameters where the phase transition into the spiral spin liquid does not break any symmetries and where the critical exponents do not match those of standard universality classes. We study the importance of momentum vortices in driving this phase transition and discuss the possibility of a Kosterlitz-Thouless transition of momentum vortices. Finally, we explore the regime where the rank-2 U(1) gauge theory is valid by investigating the four-fold pinch point singularities across the phase diagram.
\end{abstract}

\date{\today}

\maketitle 

\section{Introduction}

Spin liquids (SLs) are one of the most sought-after states in the field of modern condensed matter physics~\cite{Balents10, Savary17, Broholm20}. This is because these disordered yet highly correlated spin states offer an ideal platform to study a wide range of captivating emergent phenomena~\cite{Bramwell01, Castelnovo08, Fennell09, Morris09, Alzate19}. Examples include hidden topological order, long-range entanglement, fractional excitations, fracton physics, and many more. In the classical limit, where spins behave as unit vectors, much attention has been focused on lattices of corner-sharing frustrated motifs such as triangles or tetrahedra~\cite{Chalker92, Huse92, Isakov04, Moessner98, Hopkinson07}. These lattices allow for a re-writing of a nearest neighbor Heisenberg Hamiltonian into a sum of complete squares over the triangular or tetrahedral building blocks. Hence, all ground states are determined by local spin constraints such as the condition of a vanishing total spin on each of these motifs \cite{Davier23, yan2023classification, yan2023classification2, fang2023classification3}. Prominent examples occur on kagome and pyrochlore lattices, where extensively degenerate ground state manifolds are connected by local spin flips and the spin structure factors exhibit characteristic two-fold pinch-point singularities. Upon adding quantum fluctuations these highly degenerate ground states comprise a well-known route toward inducing quantum spin liquid behavior~\cite{Sachdev92, Lecheminant97, Yan11, Messio12, Gingras14}. Particularly fascinating is the possibility of realizing emergent gauge theories via these local spin constraints, which in pyrochlore quantum spin ice systems gives rise to an effective U(1) quantum electrodynamics~\cite{Huse03,Hermele04}.

Another kind of spin liquid is the spiral spin liquid induced by competition between the nearest and further neighbor couplings. Contrary to the aforementioned spin liquids which result from local spin constraints, this kind of system exhibits only a subextensively degenerate classical ground-state manifold given by a continuum of minima in reciprocal space, homotopic to a ring in two dimensions or sphere (or other 2D surfaces)  in three dimensions~\cite{Bergman07, Mulder10, Gao17, Iqbal18, Buessen18, Ghosh19, Niggemann20, Yao21, Gao22, Graham23}. In these systems, however, a small change in the spiral wave vector implies a global, coordinated change of all spins in real space, contrary to the local spin flips of kagome or pyrochlore magnets. Thus, it is not {\it a priori} clear how spiral spin liquids can explore the whole ground-state manifold. A solution to this issue was proposed very recently, identifying the momentum vortices as the effective local degrees of freedom responsible for the liquidity of the phase~\cite{Yan22}. The proliferation of these topological defects allows each small patch of spins in the system to visit different wave vectors along the spiral surface at very little energy cost.

Furthermore, a low-energy effective theory for a classical spiral spin liquid was derived, which possesses remarkable similarities with a rank-2 U(1) gauge theory. The family of such theories constitutes higher-rank generalizations of U(1) electromagnetism known for their unusual kinetically constrained {\it fracton} quasiparticles~\cite{Pretko17, Pretko17b, Yan22}. As a result of the tensor structure of the associated Gauss' law, a rank-2 U(1) gauge theory exhibits four-fold pinch-point singularities in the electric-field correlator~\cite{Prem18} instead of the usual two-fold pinch points of the electric-field correlator in standard U(1) electromagnetism. 
The effective theory also has a close relation to the fracton theory of smectic matter \cite{PhysRevA.106.023321, ZHAI2021168509, PhysRevLett.125.267601, radzihovsky2023critical, Machon2019}. In particular, the topological defects known as momentum vortices have the same exotic feature as those in smectic matter, namely the winding number cannot be larger than 1, but can be any arbitrary negative number. The mathematics behind this has been discussed in detail in Refs.~\cite{Yan22, Machon2019}.

In the spiral system of Ref.~\cite{Yan22} -- a classical square lattice XY model with up to third nearest neighbor interactions -- the emergent electric field translates into a second-order derivative of the spin angle field and the corresponding correlation function of these objects has indeed been numerically confirmed to show four-fold pinch points. Nevertheless, the range of validity of the rank-2 U(1) gauge theory emerging from a spiral spin liquid remains uncharted since the exact mapping requires the radius of the spiral contour in momentum space to be nearly vanishing. A further question left unanswered in Ref.~\cite{Yan22} is the nature of the thermal phase transition observed as a sharp heat capacity peak when the system enters the spiral spin liquid regime.

Motivated by the previous results but also by remaining open questions, in this article, we revisit the spiral phase of the classical XY model on the square lattice using classical Monte Carlo and molecular dynamics calculations. Compared to Ref.~\cite{Yan22} where the system was only studied for a single parameter set for a fixed, small, and almost circular spiral ring, here, we explore a larger region of the phase diagram with varying radii of the spiral contour, revealing the broader interplay of phases in which the spiral spin liquid is embedded. Particularly, moving away from the circular spiral ring we study the range of validity of the correspondence between the spiral spin liquid and the rank-2 U(1) gauge theory. We also uncover a rich phase diagram in temperature and parameter space that contains a variety of interesting magnetic phases beyond the previously studied spiral spin liquid, pancake liquids, and rigid vortex networks. This includes double stripe states stabilized by two intertwined entropic selection mechanisms that lead to a $\mathds{Z}_2\times \mathds{Z}_2$ symmetry breaking of lattice translation and rotation symmetries. However, since these selection mechanisms have a finite energy cost, at lower temperatures the system regains one of the broken $\mathds{Z}_2$ symmetries, showing a re-entrant behavior into a nematic spiral phase. The different broken symmetries lead to phase transitions belonging to the 2D Ising and four-state Potts universality classes. Furthermore, between the pancake and spiral spin liquids, we observe a phase transition with a logarithmically divergent specific heat that is not related to any symmetry breaking. We investigate the behavior of momentum vortices at this transition and discuss the possibility of a Kosterlitz-Thouless transition of momentum vortices.

The remainder of the article is organized as follows: Section~\ref{sec:modandmet} introduces the XY model on the square lattice and the key properties of the spiral contour. In Sec.~\ref{sec:res} we show the finite-temperature phase diagram as obtained in our calculations, with different subsections devoted to the individual phases. In Sec.~\ref{sec:dyn} we investigate the dynamical spin structure factor and discuss the characteristic spectral features of the different phases. In Sec.~\ref{sec:vortphys} we discuss the roles of spin and momentum vortices, especially at the phase transition into the spiral spin liquid regime. Finally, in Sec.~\ref{sec:pinchpoints} we study the validity of the rank-2 U(1) gauge theory throughout the phase diagram. Sec.~\ref{sec:conc} contains the conclusions of our work.

\section{spiral spin model}
\label{sec:modandmet}

\begin{figure}[!t]
\centering
\includegraphics*[width=0.35\textwidth]{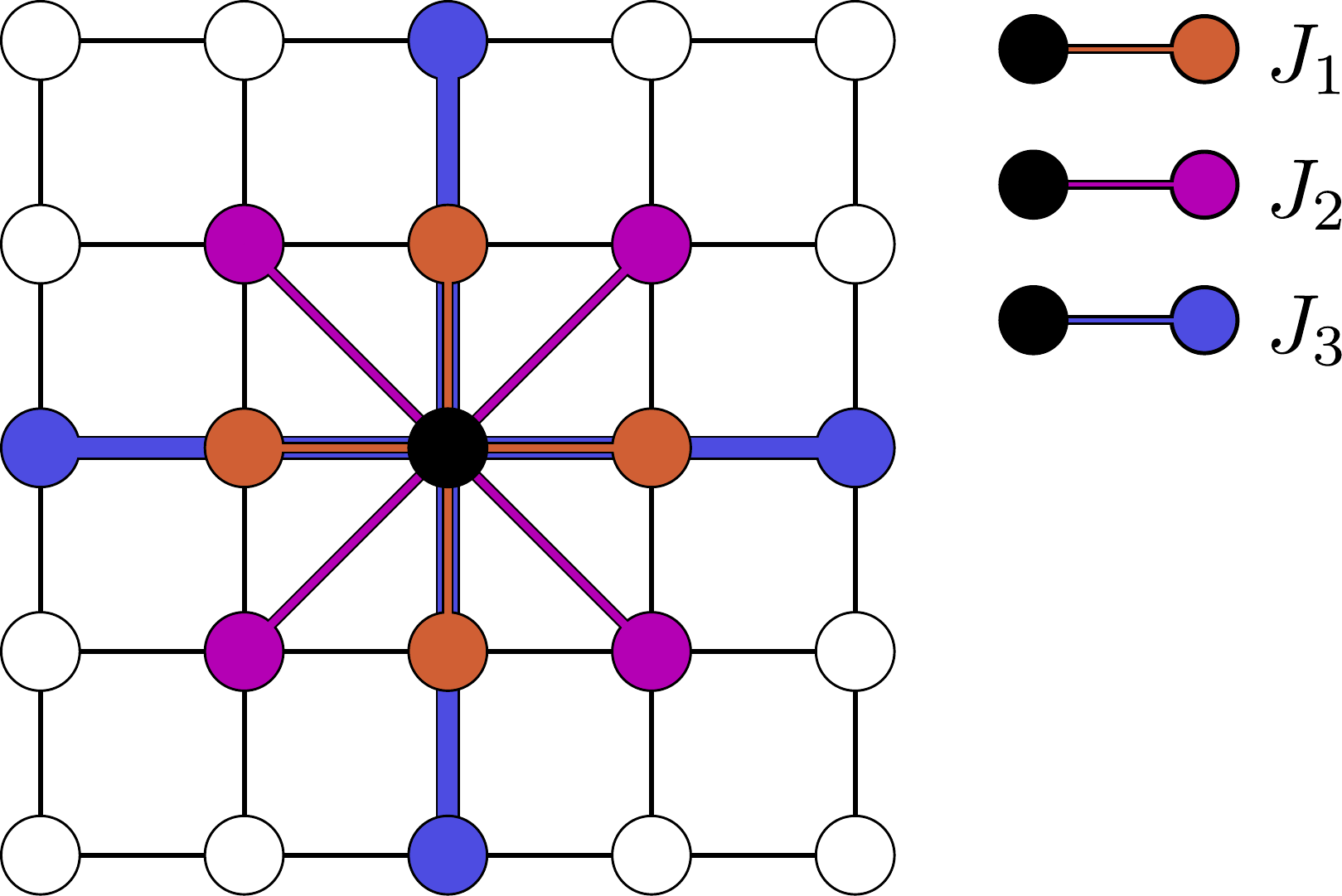}
\caption{Square lattice with up to third-nearest-neighbor exchange interactions shown in orange, purple, and blue, respectively, emanating from an arbitrary site (black).}
\label{fig:lattice}
\end{figure}

\begin{figure}[!t]
\centering
\includegraphics*[width=0.35\textwidth]{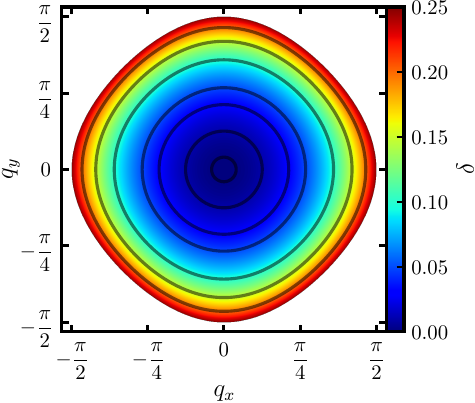}
\caption{Continuous ground-state manifolds of wave-vectors as a function of $\delta$ corresponding to the spiral solutions of Eq.~(\ref{eq:rings}). Contour lines are shown for $\delta = 0.001$, 0.01, 0.03, 0.05, 0.10, 0.15, and 0.20 from smallest to largest.}
\label{fig:spisurf}
\end{figure}

The Hamiltonian for the XY model on the square lattice considered in this article is given by
\begin{equation}
\mathcal{H} = J_1 \sum_{\langle ij\rangle_1} \mathbf{S}_i \cdot \mathbf{S}_j +  J_2 \sum_{\langle ij\rangle_2} \mathbf{S}_i \cdot \mathbf{S}_j +  J_3 \sum_{\langle ij\rangle_3} \mathbf{S}_i \cdot \mathbf{S}_j,
\label{eq:ham}
\end{equation}
where $\langle ij\rangle_1$, $\langle ij\rangle_2$, and $\langle ij\rangle_3$ indicate first-, second-, and third-nearest neighbor sites, respectively (see Fig.~\ref{fig:lattice}). The XY spins $\mathbf{S}_i=(S_{ix},S_{iy})$ are normalized as $|\mathbf{S}_i|=1$ and represented by an angle $\phi \in \left[0,2\pi\right)$ via $\mathbf{S}_i=(\cos \phi_i, \sin \phi_i)$. Throughout this work we constrain the coupling parameters as follows: We fix the nearest-neighbor interaction to be ferromagnetic, $J_1=-1$, and consider antiferromagnetic $J_2$ and $J_3$. Furthermore, we fix $J_2 > 0.25$ and $J_3 = J_2/2$, as this gives rise to the spiral ground states of interest here~\cite{Yan22}. It is thus convenient to use the parametrization $\delta = J_2-0.25$, such that the transition between a pure ferromagnet and spin spirals occurs at $\delta = 0$. The degenerate momenta $\mathbf{q}=(q_x, q_y)$ of the spin spirals lie on a ring in reciprocal space, determined by the solutions of the equation
\begin{equation}
\cos(q_x) + \cos(q_y) = \frac{2}{4\delta+1}.
\label{eq:rings}
\end{equation}
In Fig.~\ref{fig:spisurf} we show the continuous manifolds of ground state wave-vectors in reciprocal space for varying values of $\delta$ from $\delta= 0$ to $0.25$. When $\delta=0.25$ the spiral ring reaches the special wave-vectors $(\pm \frac{\pi}{2}, 0)$ and $(0, \pm \frac{\pi}{2})$.

From Eq.~(\ref{eq:rings}) it is clear that $\delta = 0$ only allows for the ferromagnetic solution $\mathbf{q}=(q_x, q_y)=(0,0)$. When $\delta$ is finite but $\delta\ll1$, solutions in the vicinity of $\mathbf{q}=(0,0)$ can be found by approximating $\cos(q) = 1 - \frac{q^2}{2} + \mathcal{O}(q^4)$. Inserting this into Eq.~(\ref{eq:rings}) results in circular rings given by
\begin{equation}
\left( \frac{q_x}{4\sqrt{\delta}}\right)^2 + \left( \frac{q_y}{4\sqrt{\delta}}\right)^2 = 1,
\end{equation}
where the radius is $4\sqrt{\delta}$~\cite{Yan22}. This is evidenced in Fig.~\ref{fig:spisurf}, where contour lines with small $\delta$ close to the origin become more circular. It is in this limit that the spiral spin liquid physics is governed by momentum vortices whose effective theory corresponds to a rank-2 U(1) gauge theory~\cite{Yan22}.

\section{Phase Diagram}
\label{sec:res}

We perform classical Monte Carlo (cMC) calculations in a range of the phase diagram given by $\delta \in [0,0.25]$, from the ferromagnetic limit to the point where the spiral ring touches the wave vectors $(\pm \frac{\pi}{2}, 0)$ and $(0,\pm \frac{\pi}{2})$. Note that the last wave vectors correspond to spirals with a periodicity of four sites along one lattice direction. The behavior in this case is expected to be substantially different from the spiral liquid studied in Ref.~\cite{Yan22}. Our core calculations are performed with the following specifications. We consider square systems with periodic boundary conditions, with up to $N=200\times 200= 40000$ sites. A logarithmic cool-down protocol is implemented with 120 temperature steps from $T=2$ down to $T=0.02$ (where the units of energy are eliminated via $|J_1| = 1$). Each temperature step consists of $5\times 10^5$ Monte Carlo steps composed of $N$ Metropolis trials and $N$ over-relaxation steps each. The acceptance ratio remains close to $50\%$ thanks to the adaptive Gaussian step~\cite{Alzate19}. Data for $e$ and $c_v$ is collected after performing half of the cMC steps at a given temperature. Results are averaged over 10 independent runs. Apart from these core calculations, we perform additional calculations, e.g. for the spin structure factors, starting from stored thermalized states.

\begin{figure}[!t]
\centering
\includegraphics*[width=0.45\textwidth]{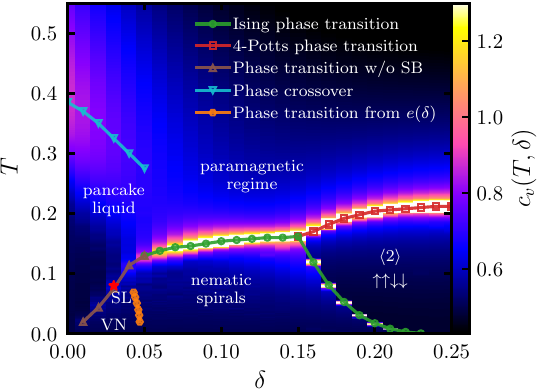}
\caption{Phase diagram of the Hamiltonian in Eq.~(\ref{eq:ham}) obtained by cMC. The color indicates the size of $c_v$. Open symbols and lines indicate the positions of the peaks or local maxima in $c_v$ (see main text), except for the orange symbols that are calculated from an energy crossover. The type of phase transition (for example, the universality class) is indicated in the legend (SB = symmetry breaking), and the phases are indicated in the corresponding regions (SL = spiral liquid, VN = vortex network). The red star corresponds to the transition into the spiral liquid as obtained in Ref.~\cite{Yan22}. }
\label{fig:phased}
\end{figure}

The energy and specific heat calculations exhibit finite-temperature phase transitions in the full range of investigated $\delta>0$. The phase diagram is presented in Fig.~\ref{fig:phased}, where the background color indicates the magnitude of the specific heat $c_v(T,\delta)$. The colors of the lines and symbols indicate the types of phase transitions manifested by peaks in the specific heat, which we will explain in detail in the following subsections. Overall, the phase diagram reveals many interesting features such as the existence of two phase transitions (red and green lines) at $\delta\geq 0.16$, where the lower critical temperature $T_c$ goes to zero as $\delta \to 0.25$. These transitions surround a double-striped phase which we denote $\langle 2 \rangle$. The central part of the phase diagram is dominated by incommensurate lattice nematic spirals which can be understood as stripy spin states in one of the two Cartesian directions. At small $\delta\lesssim 0.05$, the system shows a crossover at high temperatures from the paramagnetic regime into a pancake liquid phase~\cite{Shimokawa19}. Upon lowering the temperature further, the system undergoes a transition into a spiral spin liquid phase (brown line) without displaying any signs of symmetry breaking. The red star on this line highlights the phase transition found in Ref.~\cite{Yan22}. At even lower temperatures, this phase turns into a rigid vortex network. 

These phase transitions and crossovers depict a rich phase diagram including the spiral spin liquid phase at finite temperatures, which we will characterize in the following. Before we continue describing the individual phases in detail, it is important to note that the Mermin-Wagner theorem forbids the spontaneous breaking of the continuous U(1) spin rotation symmetry at any finite temperature. Therefore, $\langle \mathbf{S}_i \rangle = 0$ holds for any site $i$ at $T>0$. This means that whenever we speak of `symmetry breaking order' we refer to orders that manifest in spin-rotation invariant order parameters (such as spin correlations $\langle \mathbf{S}_i \mathbf{S}_j\rangle$) and that break discrete lattice symmetries.

\subsection{Double colinear stripes}
\label{sec:double}

\begin{figure}[!t]
\centering
\includegraphics*[width=0.47\textwidth]{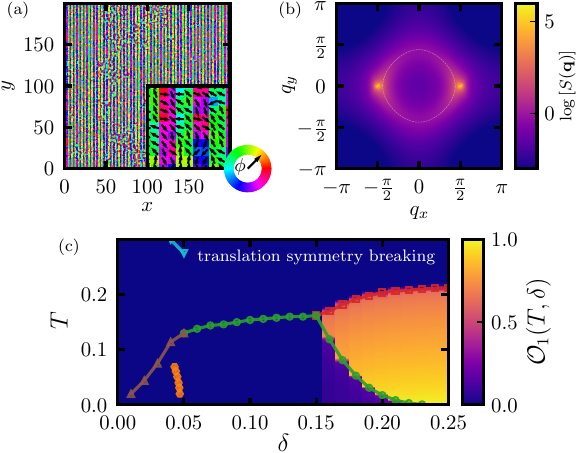}
\caption{Characterization of phase $\langle 2 \rangle$: (a) Snapshot of a real-space spin configuration from cMC for $L=200$ at $\delta = 0.17$ and $T=0.119$ (the inset shows a zoom-in), where colors indicate the angle $\phi_i$ at each site $i$. (b) Spin structure factor $S(\mathbf{q})$ (on a logarithmic scale) for $\delta = 0.17$ and $T=0.119$. (c) Order parameter $\mathcal{O}_1$ to test for lattice translation symmetry breaking [see main text and Eq.~(\ref{eq:o1})].}
\label{fig:phase2}
\end{figure}

Let us start with the parameter regime corresponding to the $\langle 2 \rangle$ phase (2-up-2-down phase), which consists of nematic two-site-wide stripes of antiparallel spins [see Fig.~\ref{fig:phase2}~(a)]. This phase breaks both lattice translation and rotation symmetries and is typically observed in systems with short-ranged ferromagnetic and long-ranged antiferromagnetic interactions~\cite{Fisher80, Seifert19, Blesio23}. In the spin structure factor, 
\begin{equation}
    S(\mathbf{q}) = \frac{1}{N} \sum_{i,j}\ \langle \mathbf{S}_i \cdot \mathbf{S}_j \rangle \ e^{i\mathbf{q}\cdot\mathbf{r}_{ij}}
\end{equation}
where $\mathbf{r}_{ij} = \mathbf{r}_j-\mathbf{r}_i$ is the distance between spins $\mathbf{S}_i$ and $\mathbf{S}_j$, the $\langle 2 \rangle$ phase is characterized by peaks at ${\mathbf{q}}=(\pm \pi/2,0)$ or $(0,\pm \pi/2)$  [see Fig.~\ref{fig:phase2}~(b)]. It is important to note that these points only belong to the spiral ground-state manifold at exactly $\delta = 0.25$. For $\delta < 0.25$ there is another transition at lower temperatures that takes the system to the ground-state manifold.

The selection of {\bf q}-points on the Cartesian axes in reciprocal space is entropic and happens due to the discrete nature of the lattice. Naively, breaking the discrete $\mathds{Z}_2$ lattice rotation symmetry that transforms the wave vectors ${\mathbf{q}}=(\pm \pi/2,0)$ and $(0,\pm \pi/2)$ into each other would lead to an Ising transition, as in the well-known $J_1$-$J_2$ square lattice Heisenberg model for $J_2 > 0.5~J_1$, where simple antiferromagnetic stripes are selected~\cite{Chandra90, Weber03}. However, this is not the case here since there is an extra accidental degeneracy that is responsible for the formation of the $\langle 2 \rangle$ phase. Specifically, the Hamiltonian in Eq.~(\ref{eq:ham}) implies that all coplanar spin stripes with the periodic four-site pattern $[\phi_A \phi_B \overline{\phi}_A \overline{\phi}_B]$ along one of the two Cartesian axis have the same energy regardless of the values of $\phi_A$ and $\phi_B$ ($\phi_A$ and $\phi_B$ indicate two different angles and $\overline{\phi}_A=\phi_A+\pi$ is the opposite direction of $\phi_A$). Within this manifold there are spin spirals with a period of four lattice sites and spatially homogeneous $\pi/2$ rotation angles between neighboring spins (i.e., $\phi_A$ and $\phi_B$ differ by $\pi/2$) as well as $\langle 2 \rangle$ phases when $\phi_A=\phi_B$ or $\phi_A=\overline{\phi}_B$. The entropic term of the free energy selects the $\langle 2\rangle$ states, indicating a preference for colinear states. The two possibilities $\uparrow \downarrow \downarrow \uparrow \uparrow \downarrow  \downarrow \uparrow \cdots$ or $\downarrow \downarrow \uparrow \uparrow \downarrow \downarrow \uparrow \uparrow \cdots$ of breaking the lattice translation symmetry represent an extra $\mathds{Z}_2$ broken symmetry. Typically, there is no reason why the two $\mathds{Z}_2$ symmetries should be broken at the same temperature. However, we observe only one phase transition belonging to the 4-state Potts universality class~\cite{Wu82}, confirmed by finite-size scaling from cMC calculations (see Appendix~\ref{app:fss}). This indicates a strong order parameter coupling, leading to a broken $\mathds{Z}_2 \times \mathds{Z}_2 = \mathds{Z}_4$ symmetry. Such an emergent 4-state Potts transition through order-by-disorder was already observed in the $J_1$-$J_3$ kagome lattice Heisenberg model but as a result of a true four-fold symmetry and not a combination of two coupled two-fold symmetries~\cite{Grison20}. One possible reason for the merging of the two types of symmetry breaking is that the two order parameters are fundamentally dependent on each other: Translation symmetry breaking along a lattice direction can only occur if nematic symmetry breaking has selected this lattice direction.

As already mentioned, the phase $\langle2\rangle$ breaks translation symmetry by one lattice spacing. As we will see later, other phases of our system (such as the nematic spiral) also break translation symmetry but remain invariant under combined translation and spin rotation. What is special about the $\langle2\rangle$ phase is that it also breaks all possible combinations of translation symmetry by one lattice spacing and global spin rotations. To explicitly prove this it is useful to consider the following order parameter
\begin{equation}
\mathcal{O}_1^x=\frac{1}{N} \left| \sum_{x,y}  (-1)^x \langle \mathbf{S}_{x,y}\cdot \mathbf{S}_{x+1,y}  \rangle \right|,
\end{equation}
where we have slightly changed the notation in the sense that the subscripts $x$, $y$ in $\mathbf{S}_{x,y}$ now indicate the Cartesian coordinates of the square lattice sites $i$. This order parameter takes into account all nearest neighbor correlations in the $x$ direction with alternating signs. Analogously, $\mathcal{O}_1^y$ can be defined that detects translation symmetry breaking by one lattice spacing in the $y$-direction. To take into account both types of translation symmetry breaking along the $x$ and $y$ axes we consider the order parameter
\begin{equation}
\mathcal{O}_1 = \mathcal{O}_1^x + \mathcal{O}_1^y.
\label{eq:o1}
\end{equation}
It is easy to see that for an ideal $\langle2\rangle$ state (with either stripes along the $x$ or $y$ directions), $\mathcal{O}_1 = 1$. In the same manner, one can check that $\mathcal{O}_1 = 0$ for ferromagnetic, antiferromagnetic N\'eel, and spiral states with homogeneous rotation angles. 

The results for $\mathcal{O}_1$ as a function of $T$ and $\delta$ are shown in Fig.~\ref{fig:phase2}~(c). The order parameter $\mathcal{O}_1$ takes a finite value only in the region encapsulated by the green and red curves at high values of $\delta$, that has been denoted $\langle2\rangle$ in Fig.~\ref{fig:phased}. Even though favored by thermal fluctuations, the $\langle 2 \rangle$ phase (and more generally, all $[\phi_A \phi_B \overline{\phi}_A \overline{\phi}_B]$ states) is not a part of the ground-state manifold for $\delta < 0.25$. Therefore, we cannot strictly speak of order by disorder due to thermal fluctuations. As a consequence, another phase transition must occur at lower temperatures for $\delta < 0.25$, taking the system to the ground-state manifold. This is confirmed by the recovery of the translation symmetry at lower temperatures (below the green line), where $\mathcal{O}_1$ takes again small values, evidencing a re-entrance phenomenon. 

We note that there is still some reminiscence of the $\langle 2 \rangle$ phase after the system passes through it at lower temperatures, manifested by the non-zero value of $\mathcal{O}_1$ below the green line at high values of $\delta$. The boundary conditions are responsible for this spurious effect, as the system tries to evolve from a 4-spin periodic structure ($\langle 2 \rangle$ phase) onto an incommensurate spiral phase. The frustration induced by the boundary conditions may forbid the system from fully re-arranging into a nematic spiral. Another consequence of this is the larger energy difference between these states when we reach $T=0$ and the exact ground-state energy
(see Appendix~\ref{app:gse}).

\subsection{Nematic coplanar spirals}
\label{sec:nematic}

\begin{figure}[!t]
\centering
\includegraphics*[width=0.47\textwidth]{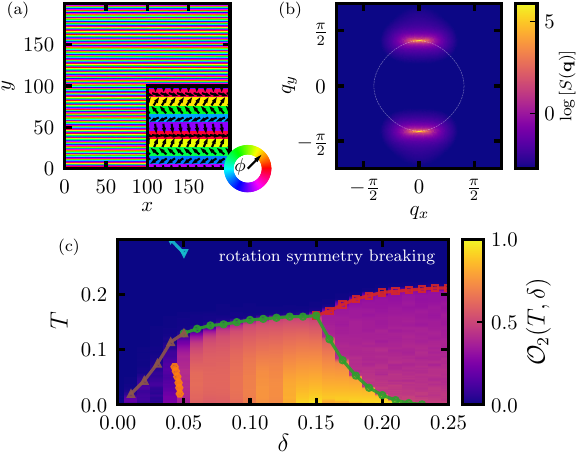}
\caption{Characterization of the nematic spiral phase: (a) Real-space configuration for $L=200$ at $\delta = 0.14$ and $T=0.0001$ (the inset shows a zoom-in), where colors indicate the angle $\phi_i$ at each site $i$. (b) Spin structure factor (in logarithmic scale) for $\delta = 0.14$ and $T=0.02$. (c) Order parameter $\mathcal{O}_2$ to test for lattice rotation symmetry breaking (see main text).}
\label{fig:phasenem}
\end{figure}

The nematic spiral phase exists below the green curve in the phase diagram of Fig.~\ref{fig:phased} and only breaks the lattice rotation symmetry by choosing two equivalent points on the spiral ring [see Fig.~\ref{fig:phasenem} (a) and (b)]. The spiral states in this phase are characterized by spatially homogeneous rotation angles between spins on neighboring sites. For $\delta \ge 0.16$, the nematic spiral phase is reached through the recovery of the translation symmetry that is broken in the $\langle 2 \rangle$ phase. The corresponding transition is found to belong to the Ising universality class (see Appendix~\ref{app:fss}) which becomes evident when approaching the phase boundary from the low-temperature side. Below the phase transition, the system only breaks $\pi/2$ lattice rotation symmetry while translation symmetry (in combination with a properly chosen spin rotation) is intact. On the other hand, above the phase transition (i.e. in the $\langle2\rangle$ phase), an extra $\mathds{Z}_2$ translation symmetry is broken, explaining the Ising nature of the transition.

For $0.05\lesssim\delta\leq0.15$, the nematic spiral phase is directly reached from the paramagnetic regime, populating a large region of the phase diagram. In this $\delta$ region, the transition is also in the Ising universality class due to the broken $\mathds{Z}_2$ lattice rotation symmetry. For most values of $\delta$, these spirals are incommensurate since the wavelength changes smoothly with $\delta$. For example, commensurate spirals with wavelengths of 4, 5, and 6 lattice spacings are realized for values $\delta$ given by 0.25, $\sim0.131$, and $\sim0.083$, respectively [see Eq.~(\ref{eq:rings})]. As a consequence, finite lattices have an extra degree of frustration due to the boundary conditions which lead to deviations from the ideal wave vectors $(q_x, 0)$ and $(0, q_y)$. These effects, however, vanish in the thermodynamic limit where order-by-disorder effects can select the exact wave vectors $(q_x, 0)$ and $(0, q_y)$ on the spiral ring.

An order parameter that detects the lattice rotation symmetry breaking can be constructed in terms of the local momentum (i.e. local wave vector) of the spiral. At each site, we can define the local momentum via $\mathbf{q} = \nabla \phi$~\cite{Yan22} (where $\nabla$ is implemented as a discrete derivative on the square lattice) which is a two-component vector in the $xy$-plane. Then, our order parameter for lattice rotation symmetry breaking is defined by
\begin{equation}
\mathcal{O}_2 = \frac{1}{N} \left| \sum_{i \in N}   \langle  \left| \hat{\mathbf{q}}_{i}^x \right| \rangle - \langle  \left| \hat{\mathbf{q}}_{i}^y \right| \rangle + \langle  \left| \hat{\mathbf{q}}_{i}^{x+y} \right| \rangle - \langle  \left| \hat{\mathbf{q}}_{i}^{x-y} \right| \rangle \right|,
\end{equation}
where $\hat{\mathbf{q}}_{i} = \mathbf{q}_i/|\mathbf{q}_i|$ is the normalized unit vector of $\mathbf{q}_i$ at site $i$ of the square lattice. Furthermore, the superscripts indicate the components of $\hat{\mathbf{q}}_{i}$ where $x+y$ and $x-y$ are symbolic notations for the two diagonal lattice directions. Even though these two directions are not favored by entropy, we include them to account for canted stripes that appear due to finite-size effects. In practice, this makes the results slightly smoother at small $\delta$. The $\mathcal{O}_2$ order parameter vanishes when there is no imbalance between the total momentum of the spirals in the $x$ and $y$ directions (or the $x+y$ and $x-y$ directions). On the other hand, $\mathcal{O}_2 = 1$ if a spiral is aligned along a Cartesian or a diagonal direction. 

The results for $\mathcal{O}_2$ are shown in Fig.~\ref{fig:phasenem} (c). We find that $\mathcal{O}_2$ is always sizeable below the green curve. This confirms that the peak in the heat capacity at $0.05\lesssim\delta\lesssim0.15$, identified as an Ising transition, is due to lattice rotation symmetry breaking. The small but non-zero signal observed at $\delta \lesssim 0.05$ can be attributed to the phase's slow dynamics (or bad thermalization) and a small number of independent runs trapped in a nematic state near the orange boundary. The $\mathcal{O}_2$ order parameter provides evidence of the existence of the phase transition shown in orange, which is not captured by the specific heat and which we will discuss in detail later. It is also important to note that the poor signal obtained in the $\langle 2 \rangle$ region originates from the discrete definition of the momentum. Due to the colinear double-stripe nature of the phase, the relative angles $\phi_i-\phi_j$ between neighboring sites $i$ and $j$ vary vastly on the scale of one lattice distance, making discrete derivatives ill-defined. Still, the non-zero values of $\mathcal{O}_2$ confirm that the $\langle 2 \rangle$ phase also breaks the lattice rotation symmetry.

\subsection{Pancake and spiral liquids}
\label{sec:panspi}

\begin{figure}[!t]
\centering
\includegraphics*[width=0.47\textwidth]{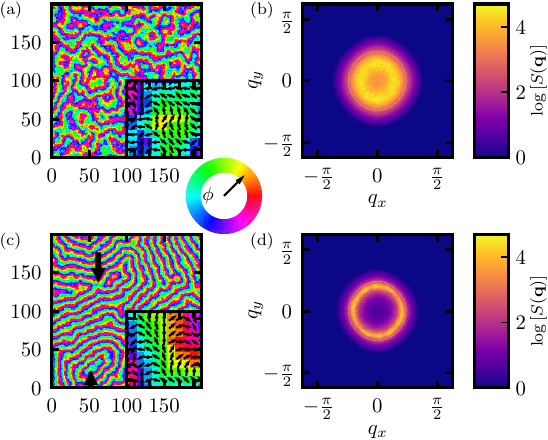}
\caption{Pancake liquid phase: (a) Real-space configuration for $L=200$ at $\delta = 0.04$ and $T=0.119$ (the inset shows a zoom-in), where colors indicate the angle $\phi_i$ at each site $i$, and (b) the corresponding spin structure factor (in logarithmic scale). Spiral liquid phase: (c) Real-space configuration for $\delta=0.04$ and $T=0.081$ where one momentum vortex and antivortex are indicated by an up and down arrow, respectively, and (d) corresponding spin structure factor.}
\label{fig:phaseliq}
\end{figure}

Finally, for $\delta\lesssim 0.06$, the heat capacity peak indicating a finite-temperature phase transition splits into two, a wide shoulder at higher $T$ signaling a crossover and a sharp peak at lower $T$ that gets smaller and shifts to $T= 0$ as $\delta$ decreases (brown curve in Fig.~\ref{fig:phased}). The broad shoulder, shown by the light-blue curve in Fig.~\ref{fig:phased}, is only discernible for $\delta \leq 0.05$ and does not exhibit critical behavior (no scaling with system size) indicating a crossover. Generally, this part of the phase diagram is well represented by the results for $\delta = 0.03$ in Ref.~\cite{Yan22}. The onset of spin correlations at the broad high-temperature peak leads to a phase known as pancake liquid~\cite{Shimokawa19}, where the spin states have contributions from spiral wave vectors ranging from $\mathbf{q}=0$ up to the spiral ring [see Fig.~\ref{fig:phaseliq} (a) and (b)]. This property leads to a spin structure factor featuring an almost uniform disk-like signal (pancake shape) within the spiral surface.

Further decreasing the temperature leads to a phase transition into the spiral spin liquid phase [see Fig.~\ref{fig:phaseliq} (c) and (d)], which does not show any symmetry breaking. Contrary to the pancake liquid, this phase presents a nearly isotropic signal in the spin structure factor only along a circle inside the spiral ring [highlighted by dashed white lines in Fig.~\ref{fig:phaseliq} (d)]. This phase shows liquid-like fluctuations between spin spirals with all possible directions of spiral wave vectors, justifying the name {\it spiral spin liquid}. These fluctuations give rise to well-defined momentum vortices and antivortices representing local defects in the spiral configurations~\cite{Yan22} [see up and down arrows in Fig.~\ref{fig:phaseliq} (c)]. The circle formed in the spin structure factor slightly enlarges as the temperature decreases until it reaches the spiral surface corresponding to the ground-state manifold. At the same time, the dynamics of the phase are slowed down and thermalization becomes difficult in the cMC runs. The transition separating the pancake and spiral spin liquids has a logarithmically divergent specific heat (see Appendix~\ref{app:fss}), corresponding to a critical exponent $\alpha=0$. Typically, $\alpha = 0$ could be associated with an Ising transition, as it is found for $\delta > 0.05$. However, in this case, the critical exponent for the correlation length $\nu$ is not consistent with an Ising universality class, and the spiral spin liquid does not exhibit any signs of symmetry breaking. This evidences that the transition falls out of the standard paradigm of phase transitions, and we will argue in Sec.~\ref{sec:vortphys} that it is possibly a Kosterlitz-Thouless transition driven by momentum vortices.

\begin{figure}[!t]
\centering
\includegraphics*[width=0.45\textwidth]{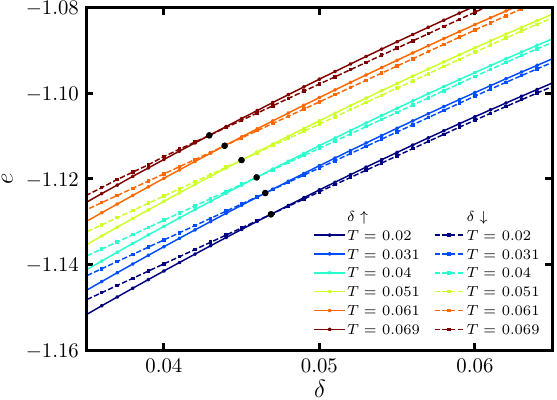}
\caption{Energy-level crossings between the nematic spiral and spiral liquid/vortex network phases. Dashed lines indicate decreasing $\delta$-sweeps starting from $\delta = 0.07$. Full lines indicate increasing $\delta$-sweeps starting from $\delta = 0.03$. Black dots indicate the crossing points.}
\label{fig:enercross}
\end{figure}

As mentioned before, two different phases are observed at low temperatures around $\delta \approx 0.05$. On one hand, we find nematic spirals for $\delta \gtrsim 0.05$, which break the lattice rotation symmetry (green curve in Fig.~\ref{fig:phased}). On the other hand, we find a spiral liquid for $\delta \lesssim 0.05$, which does not break any symmetries (brown curve in Fig.~\ref{fig:phased}). Thus, a phase transition \textit{must} exist between these two phases. However, no such phase transition is captured by the specific heat, even when very small $\delta$ steps are considered (see Appendix~\ref{app:zoomin}). Furthermore, performing $\delta$-sweeps at constant $T$, coming from either of the two phases, we could not tune the system from one phase into the other. This can be attributed to the large change of spin configuration needed to go from one phase to the other, the slow dynamics of spin fluctuations at these temperatures, and/or the frustration generated by the boundary conditions. Other update methods for cMC could provide a solution to this challenging problem which, however, we leave for future work.  

Despite the difficulties in evolving the two phases into each other, we can infer the existence of a phase transition from energy considerations. Specifically, Fig.~\ref{fig:enercross} shows an energy-level crossing in our $\delta$-sweeps at constant $T$, when either starting in the nematic spiral phase and decreasing $\delta$ (dashed lines) or starting in the spiral liquid/vortex network phase and increasing $\delta$ (full lines). The crossing points are denoted by black dots and are shown by orange symbols in Fig.~\ref{fig:phased}. This line roughly agrees with the region in the phase diagram where the order parameter $\mathcal{O}_2$, detecting lattice rotation symmetry breaking, vanishes. Nonetheless, it is worth noticing that this method of locating the phase transition is only strictly valid at $T=0$ or when the two phases have the same entropy (which is probably not the case here). Therefore, the orange line in Fig.~\ref{fig:phased} should only be understood as a rough estimate for the transition. Since we could not observe the system's evolution from one phase to the other, we could also not detect any hysteresis behavior.

\subsection{Vortex network or ripple state}
\label{sec:vortexnet}

\begin{figure}[!t]
\centering
\includegraphics*[width=0.47\textwidth]{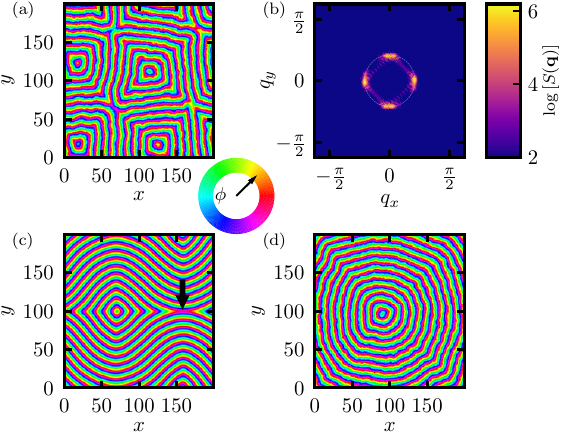}
\caption{Vortex network phase: (a) Real-space configuration for $L=200$ at $\delta = 0.03$ and $T=0.02$, and (b) the corresponding spin structure factor. (c) Real-space configuration for $\delta=0.03$ and $T=0.0001$, where the black arrow shows an elongated momentum antivortex. (d) Ripple state obtained for $\delta= 0.03$ and $T=0.02$ with open boundary conditions.}
\label{fig:phasevn}
\end{figure}

As mentioned in the previous section, the ring-like signal in the spin structure factor characterizing the spiral liquid (see Fig.~\ref{fig:phaseliq}) grows as the temperature is lowered until it reaches the spiral ring of the exact ground state manifold. Around this temperature, a crossover to a rigid vortex network state occurs, where the spirals are well-defined and correspond to four different wave vectors $\mathbf{q}=(\pm q, 0)$ and $(0,\pm q)$~\cite{Yan22} selected by an order-by-disorder mechanism. A typical spin configuration is shown in Fig.~\ref{fig:phasevn} (a), with the corresponding spin structure factor presented in Fig.~\ref{fig:phasevn} (b). This rigid vortex network does not show any indications of symmetry breaking and realizes an approximate square arrangement of momentum vortices and antivortices (therefore the name vortex network) connected by straight domain walls~\cite{Yan22}. Furthermore, these structures are rigid, in the sense that they evolve very slowly in Monte Carlo time. This makes the system very prone to getting stuck in a metastable spin configuration. For example, while the wave vectors $\mathbf{q}=(\pm q, 0)$ and $(0,\pm q)$ are favored by entropic effects at finite temperature, we still find configurations with $\mathbf{q}=\pm (q, q)$ and $\pm(q,-q)$, indicating that the cMC runs are affected by thermalization issues, as well as by finite sizes and boundary conditions.

Other studies of a closely related model on the honeycomb lattice found that the low-temperature phase is a ripple state for systems with open boundary conditions~\cite{Shimokawa19}. The ripple state consists of a single momentum vortex from which spin spirals extend in all directions. Interestingly, we also find such a state in the vortex network regime when imposing open boundary conditions [see Fig.~\ref{fig:phasevn} (d)]. The discrepancy between the two boundary conditions indicates that they play an essential role in determining the bulk spin configurations even for large lattice sizes. This applies in particular to the limit $\delta \to 0$ where the spiral wavelength diverges such that it is not possible to make a sensible extrapolation to the thermodynamic limit. Nonetheless, at finite temperatures, a ripple state in the thermodynamic limit seems peculiar because this would imply that spins far away from the ripple center are affected by it, regardless of the distance. In what follows, we provide a simple understanding of why the single vortex state is favored by the open boundary condition.

\begin{figure}[!t]
\centering
\includegraphics*[width=0.42\textwidth]{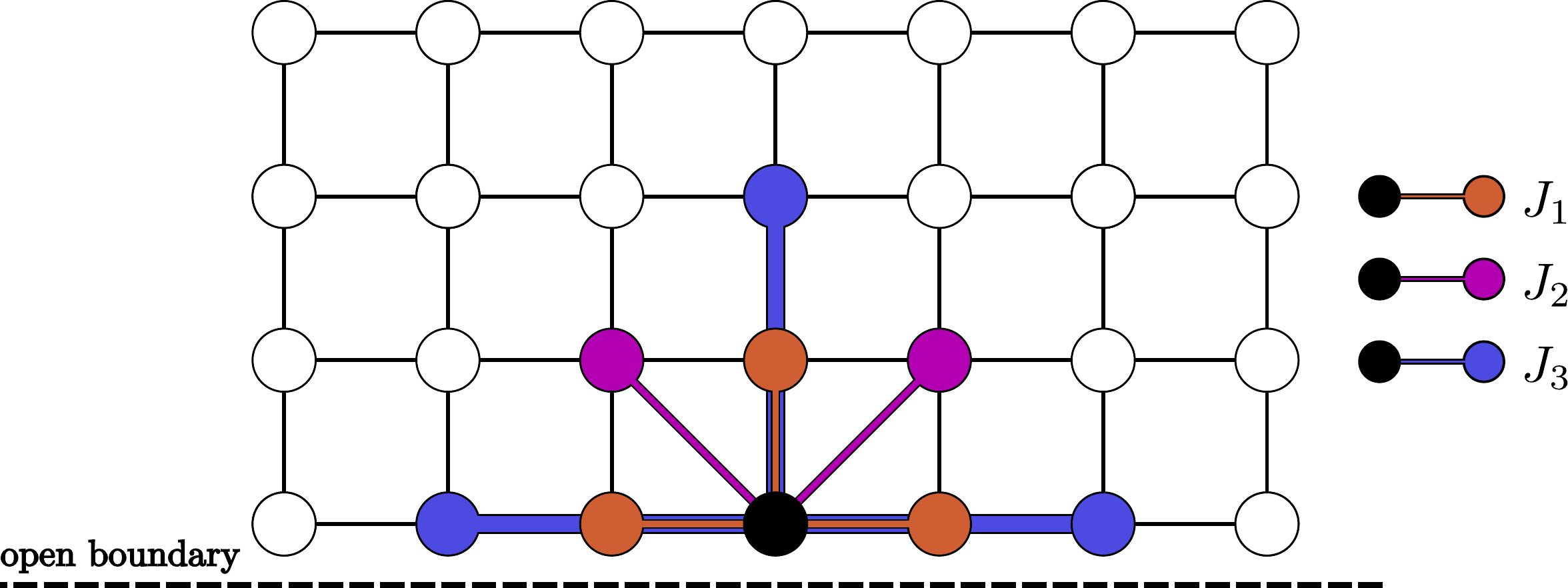}
\caption{Bottom part of a lattice with an open boundary and the remaining couplings to the system of a spin at the edge (black circle).}
\label{fig:latopen}
\end{figure}

As an example, we examine the bottom boundary of a square-shaped system (i.e., a boundary oriented along the $x$-axis) and investigate the Hamiltonian at the boundary. The other three boundaries of the system can be treated analogously. Assuming that the spins are in a spiral state with wave vector $\mathbf{q} = (q_x, q_y)$, the energy of a spin closest to the boundary can be obtained by computing the Hamiltonian in Eq.~(\ref{eq:ham}) for the spiral state but omitting the contributions from the bonds that are cut by the boundary. This leads to an energy contribution from such a spin given by (see Fig.~\ref{fig:latopen})
\begin{equation}
\begin{split}
 2 e_\text{bdry}  =   &    2J_1 \cos{(q_x a )} + J_1 \cos{(q_y a) } \\
   & + J_2 \cos{(q_x a + q_y a)} + J_2 \cos{(-q_x a + q_y a)} \\
&+ 2J_3  \cos{(2 q_x a) } + J_3  \cos{ (2 q_y a )},
\end{split}
\end{equation}
where we recall that $J_1= -1$, $J_2 = 1/4+\delta$, $J_3 = 1/8+\delta/2$ and $a=1$ is the nearest-neighbor distance. Minimizing $e_\text{bdry}$ as a function of $q_x$ and $q_y$  leads to a wave vector $\mathbf{q}' = (0, \pm q_y')$ with $q_y'$ given by
\begin{equation}
\cos(q_y' a) = \frac{1 - 4 \delta}{ 1 + 4 \delta}.
\end{equation}
Note that since $\cos(0) + \cos(q_y' a ) =\frac{2}{ 1 + 4 \delta} $, the wave vector $(0, \pm q_y')$ is exactly on the spiral ring [see Eq.~(\ref{eq:rings})].

This analysis shows that open boundaries energetically prefer spirals whose stripe-like spin patterns are parallel to the boundary, as is (approximately) realized for the ripple state in Fig.~\ref{fig:phasevn}   (d). While the bulk energy is minimized by a domain with a single wave vector $\mathbf{q}$ on the spiral ring, such a state costs much energy on the boundary (growing linearly with the linear system size $L$) since a single domain cannot simultaneously satisfy the energetic preferences at all boundaries. Hence, the system forms a single momentum vortex in the bulk which reduces the energy costs at the boundary (note that the bulk energy of a single momentum vortex only scales logarithmically in the system size $L$~\cite{Yan22}). This mechanism stabilizes the single vortex ripple state in Fig.~\ref{fig:phasevn} (d) at low temperatures when open boundary conditions are imposed.

Upon further decreasing the temperature, systems with periodical boundary conditions show a re-organization of momentum vortices associated with a bending of the spiral wavefronts. This leads to spin configurations shown in Fig.~\ref{fig:phasevn} (c) obtained at $T=0.0001$. The bending of the wavefronts gives rise to momentum antivortices with a very elongated shape along one Cartesian direction [shown by the black arrow in Fig.~\ref{fig:phasevn} (c)]. Selecting one of the two lattice directions corresponds to a $\mathds{Z}_2$ symmetry breaking and implies a certain degree of lattice nematicity in the system. This spontaneous breaking of a $\mathds{Z}_2$ symmetry is expected at low but finite temperatures. However, due to the aforementioned thermalization issues in the vortex network phase, most independent cMC runs do not present any signatures of this transition in the specific heat such that the numerical evidence for the spontaneous symmetry breaking remains rather weak. On the other hand, the decreasing density of momentum vortices at low temperatures is a very robust observation. When the distance between momentum vortices becomes comparable to the system size, the results are largely affected by the boundary conditions and therefore cannot be trusted. We thus conclude that the vortex network regime is the numerically most challenging part of the phase diagram and a thorough understanding of this phase is still lacking. 

\section{Dynamics}
\label{sec:dyn}

\begin{figure}[!t]
\centering
\includegraphics*[width=0.45\textwidth]{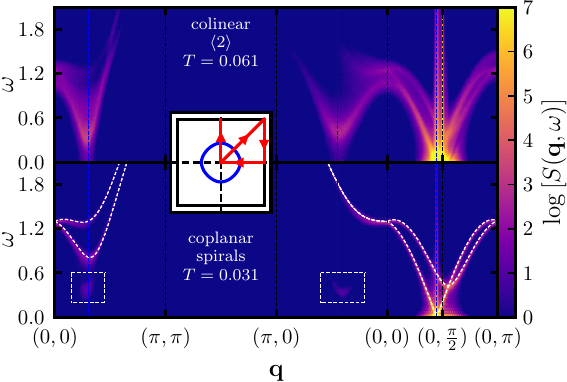}
\caption{Spin structure factor $S(\mathbf{q},\omega)$ from MD calculations at $\delta = 0.18$ for two different temperatures, corresponding to the colinear $\langle 2 \rangle$ phase (top panel) and the coplanar nematic spiral phase (bottom panel). The dashed white lines are the spin wave bands obtained by LSWT at $T=0$, and the dashed blue vertical lines indicate momenta on the spiral ring. The inset shows the plotted path in momentum-space (red) and the spiral ring corresponding to the ground-state manifold (blue). The white dashed boxes indicate spurious ghost signals (see main text).}
\label{fig:dyn1}
\end{figure}

\begin{figure}[!t]
\centering
\includegraphics*[width=0.45\textwidth]{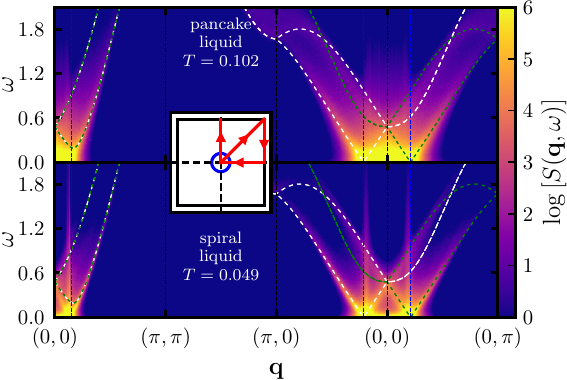}
\caption{Spin structure factor $S(\mathbf{q},\omega)$ from MD calculations at $\delta = 0.03$ for two different temperatures, corresponding to the pancake liquid (top panel) and the spiral spin liquid (bottom panel). The white and green dashed lines correspond to the spin wave bands obtained by LSWT at $T=0$ for ground state nematic spirals with wave vectors $\mathbf{q}=(q,0)$ and $\mathbf{q}=(0,q)$ on the spiral ring, respectively. The dashed blue vertical lines indicate the momenta on the ground state spiral ring. The plotted path in $\mathbf{q}$-space is shown in the inset (red), together with the spiral ring (blue). Note that the color scale is logarithmic.}
\label{fig:dyn2}
\end{figure}

To study the signatures of the different phases in the {\it dynamical} spin structure factor which is experimentally accessible through inelastic neutron scattering, we perform molecular dynamics (MD) calculations. These simulations start from XY spin configurations obtained from cMC for a set of coupling parameters at a given temperature and then calculate the system's time evolution by solving the Landau-Lifshitz equations (without a damping term)~\cite{Moessner98,Moessner98b}. The dynamical spin structure factor is then calculated as
\begin{equation}
S(\mathbf{q},\omega) = \frac{1}{N\sqrt{N_t}} \sum_{n_t =0}^{N_t} \sum_{i,j}^N\ \langle \mathbf{S}_i(t) \cdot \mathbf{S}_j(0) \rangle \ e^{i(\mathbf{q}\cdot \mathbf{r}_{ij} - \omega t)},
\end{equation}
where the time is given by $t=n_t \delta t$, for a given time step $\delta t$~\cite{Conlon09,Zhang19}. It is important to note that the Landau-Lifshitz dynamics are only well-defined for three-component spins. Hence, for the MD calculations, we use the XY Hamiltonian from Eq.~(\ref{eq:ham}) but let it act on three-component spins~\cite{Nho02}, i.e., the spin's $z$-component (which vanishes in the initial spin state from cMC) does not contribute to the energy. During the MD runs, the energy per spin is conserved up to the fifth decimal digit, indicating that the time evolution is well performed. The dynamical spin structure factors in Figs.~\ref{fig:dyn1} and \ref{fig:dyn2} are obtained by averaging over 120 independent initial configurations. We point out that the vertical high-intensity signals extending from $\omega=0$ to the upper boundary of the plotted regions are artifacts generated by the Fourier transform in time, which can be explained as follows. The mismatch between the period of spin oscillations and the simulation time generates a finite offset in the Fourier-transformed signal. Although this offset is much smaller than the height of the peak, it is visible for the momenta where the intensity is highest. The MD results are compared with the zero-temperature spin wave bands from linear spin-wave theory (LSWT), calculated with the software SpinW~\cite{Toth15}.

In Fig.~\ref{fig:dyn1} we show results for $\delta = 0.18$ at two different temperatures, above and below the low-temperature phase transition between the $\langle 2 \rangle$ phase and the nematic spiral phase. The bottom panel shows results for the nematic spiral phase, where the magnetic Bragg peak occurs at a wave vector on the spiral ring close to $\mathbf{q}=(0,\pi/2)$. Momenta on the spiral ring that lie on the plotted path are indicated by dashed blue lines. As a reference, we also plot the spin wave bands obtained with LSWT, shown as dashed white lines in Fig.~\ref{fig:dyn1}. Overall, the spectral weight obtained within MD closely follows the dispersive bands from LSWT. Another relevant feature revealed by both methods is that, away from the Bragg peak the spectrum is always gapped, even for momenta $\mathbf{q}$ on the spiral ring, which is a consequence of the XY-anisotropy. A band minimum with a small but finite gap is found close to $\mathbf{q}=(0,\pi/2)$ and is related to the proximity to the $\langle 2 \rangle$ phase at $\delta = 0.25$. Finally, it should be noted that some lower energy ``ghost'' excitations are also observed, highlighted by dashed white rectangles. These are spurious signals originating from the passing through the colinear $\langle 2 \rangle$ phase, as was already found in the translation symmetry-breaking order parameter $\mathcal{O}_1$ (see Fig.~\ref{fig:phase2}). These features should be disregarded as remnants from the $\langle 2 \rangle$ phase. 

The top panel of Fig.~\ref{fig:dyn1} corresponds to the higher-temperature colinear $\langle 2 \rangle$ phase, where the magnetic Bragg peak is located at $\mathbf{q}=(0, \pi/2)$. Note that in this case, LSWT cannot be applied because the order is unstable at $T=0$ (it does not belong to the ground-state manifold for $\delta \neq 0.25$). In addition to the gapless excitations at the Bragg peak, there are further low-energy band minima at the wave vectors where the plotted path crosses the spiral ring (indicated by dashed blue lines). In contrast to the coplanar nematic spiral phase, the spectrum is more dispersed and the weight is not concentrated along well-defined narrow excitation modes. Overall, apart from the low-energy signal around the Bragg peak, the intensity distributions in the colinear $\langle 2 \rangle$ and the coplanar spiral phases are rather different. This is a remarkable observation given that the wave vectors of the two corresponding orders are very similar.

We now turn to the other end of the phase diagram at $\delta = 0.03$, where the pancake and spiral spin liquids are found. Figure~\ref{fig:dyn2} shows the MD results for the dynamical spin structure factor for two different temperatures corresponding to both phases. The pancake liquid phase receives its name due to the nearly homogeneous contributions from all wave vectors inside the spiral ring. This property is also observed in the dynamical spin structure factor of the pancake liquid, displayed in the top panel of Fig.~\ref{fig:dyn2}. The intensity distribution shows a strong signal at low energies for all wave vectors enclosed by the spiral ring (i.e., between the dashed blue lines) without any particular Bragg peak. Here, the white and green dashed lines represent the LSWT bands at $T=0$ for wave vectors $\mathbf{q}=(q,0)$ and $\mathbf{q}=(0,q)$ on the spiral ring, respectively. In Fig.~\ref{fig:dyn2} one can appreciate that the spectral weight of the pancake liquid follows these bands while displaying a continuum of excitations in the region enclosed by them. 

When decreasing the temperature at $\delta = 0.03$, the system enters the spiral spin liquid phase where only wave vectors close to the spiral ring coexist. This implies that spin spirals are well-defined but their momentum direction can vary in real space. The dynamical spin structure factor of the spiral spin liquid in the bottom panel of Fig.~\ref{fig:dyn2} displays this property where the low energy signal is now more concentrated around the spiral ring (blue dashed lines). Also at higher energies the spectral weight is sizeable only close to the $T=0$ LSWT bands, while the region enclosed by them remains empty, representing a significant difference from the pancake liquid. Overall these features help to distinguish the two types of spin liquids by their excitation spectrum. 

\section{Effects of spin and momentum vortices}
\label{sec:vortphys}

As mentioned in Sec.~\ref{sec:panspi}, the phase transition between the pancake liquid and the spiral spin liquid shows unusual properties. It is not associated with any spontaneous symmetry breaking and its scaling behavior falls out of the standard classification of second-order phase transitions. Specifically, the specific heat is logarithmically divergent (see Appendix~\ref{app:fss}) but the critical exponent $\nu$ is not consistent with the known universality classes. Here, we revisit the nature of this phase transition and discuss whether it could be driven by vortices, either formed by spin or momentum. Since both degrees of freedom are planar quantities, a Kosterlitz-Thouless transition seems to be a natural possibility~\cite{Kosterlitz73, Kosterlitz74}.

\begin{figure}[!t]
\centering
\includegraphics*[width=0.45\textwidth]{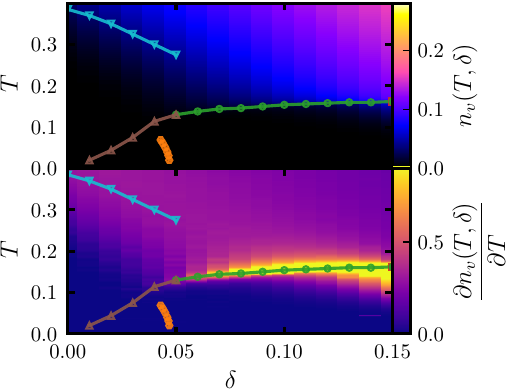}
\caption{Top panel: Spin vortex density $n_v$ (defined by the number of spin vortices per number of sites) as a function of temperature $T$ and $\delta$. Bottom panel: Derivative of $n_v$ with respect to the temperature. The lines and circles indicate the phase boundaries obtained from peaks in the specific heat.}
\label{fig:vorts}
\end{figure}

To obtain insights into whether spin or momentum vortices show any special behavior across the phase transition, we first examine the density of vortices within cMC. We start investigating spin vortices which are identified by the total spin rotation angle along elementary four-site loops of the square lattice and whose density we denote $n_v$. The results for $n_v$ are shown in the top panel of Fig.~\ref{fig:vorts} while the bottom panel displays its derivative with respect to temperature, $\partial n_v / \partial T$. In the small $\delta$-regime the spin-vortex density $n_v$ shows a pronounced drop at rather large temperatures $T\sim 0.4$. As a result of this pronounced decrease, at the low-temperature phase transition between the pancake liquid and the spiral spin liquid (brown line) $n_v$ is negligibly small. This makes it evident that the transition is not related to spin vortices and we, therefore, rule out a Kosterlitz-Thouless transition from the spin degree of freedom. Furthermore, $n_v$ increases with $\delta$, and close to the Ising transition indicated by the green line the system is still populated by a considerable number of spin vortices.

It is worth emphasizing that our investigation of the vortex density is only to check whether there is a general connection between vortices and the spiral liquid transition. The vortex density is expected to be smooth across a Kosterlitz-Thouless transition~\cite{Ota92, Lee05}, however, the derivative has been shown to exhibit a peak at the same position as that of the specific heat~\cite{Ota92}. Despite coinciding, these peaks occur at temperatures slightly above the Kosterlitz-Thouless transition and do not diverge on the square lattice XY model~\cite{Nguyen21}. Thus, $n_v$ is no good quantity to precisely locate a Kosterlitz-Thouless transition. We, nevertheless, study the vortex density here since more standard quantities for identifying a Kosterlitz-Thouless transition, such as the correlation length, are difficult to calculate accurately due to the spiral nature of the magnetic parent state (which leads to oscillating correlation functions).

\begin{figure}[!t]
\centering
\includegraphics*[width=0.45\textwidth]{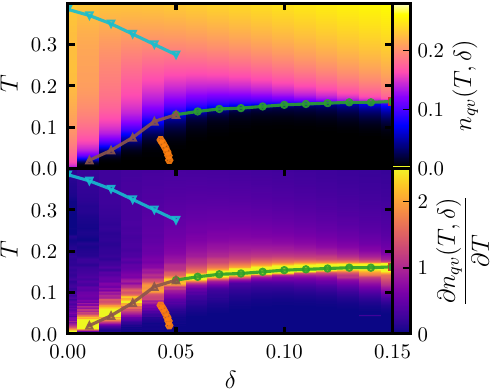}
\caption{Top panel: Momentum vortex density $n_{qv}$ (defined by the number of momentum vortices per number of sites) as a function of temperature $T$ and $\delta$. Bottom panel: Derivative of $n_{qv}$ with respect to the temperature. The lines and circles indicate the phase boundaries from the specific heat.}
\label{fig:qvorts}
\end{figure}

Next, we investigate the density of momentum vortices $n_{qv}$ in the $\delta$-$T$ phase diagram. If $\phi(\mathbf{r})$ evolves smoothly throughout the lattice, we can define the momentum field $\mathbf{q}(\mathbf{r})=\nabla \phi(\mathbf{q})$ locally as the discrete derivative of $\phi(\mathbf{r})$ and identify momentum vortices in the local momentum texture $\mathbf{q}(\mathbf{r})$. The results are shown in the top panel of Fig.~\ref{fig:qvorts} while the derivative $\partial n_{qv} /\partial T$ is presented in the bottom panel. One sees that for all values of $\delta$, the momentum vortices have a strong presence at temperatures above the phase transitions. Furthermore, $n_{qv}$ drops abruptly at the transitions as evidenced by its derivative, which is peaked along the brown line. At $\delta = 0$, where there is no finite-temperature phase transition, the momentum vortex density remains high down to the lowest temperatures. This observation indicates that momentum vortices play a key role in driving the phase transition between the pancake liquid and the spiral spin liquid.

\begin{figure}[!t]
\centering
\includegraphics*[width=0.45\textwidth]{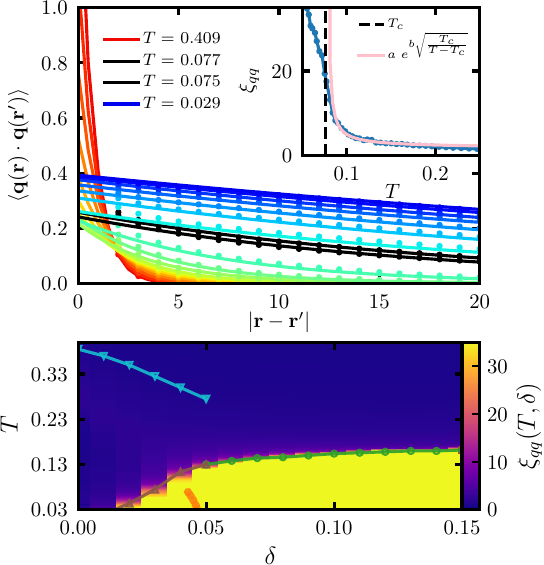}
\caption{Top panel: Decay of momentum-momentum correlations $\langle \mathbf{q}(\mathbf{r})\cdot \mathbf{q}(\mathbf{r}') \rangle$ as a function of the distance between sites $|\mathbf{r}-\mathbf{r}'|$ at $\delta=0.03$. The temperature ranges from $T=0.409$ (red) to 0.029 (blue), and lines correspond to exponential fits (black lines and points indicate data near the critical temperature). The extracted correlation lengths $\xi_{qq}$ are shown in the inset and fitted by the pink curve between $T=0.204$ and $0.084$. Bottom panel: Momentum-momentum correlation length $\xi_{qq}$ as a function of temperature $T$ and $\delta$.}
\label{fig:xi}
\end{figure}

The close connection between the spiral liquid transition and momentum vortex proliferation motivates us to analyze more closely whether this transition could be a Kosterlitz-Thouless transition of momentum vortices. To do so, we investigate the behavior of the momentum-momentum correlation length $\xi_{qq}$, which is expected to diverge at a Kosterlitz-Thouless transition $T_{\text{KT}}$ according to~\cite{Kosterlitz74, Nguyen21}
\begin{equation}
\xi_{qq}(T)\sim e^{{b\sqrt{\frac{T_\text{KT}}{T-T_\text{KT}}}}}
\end{equation} 
when approaching $T_\text{KT}$ from above. Here, $b$ is a non-universal, dimensionless number. To obtain $\xi_{qq}(T)$ we calculate the momentum-momentum correlation function $\langle \mathbf{q}(\mathbf{r})\cdot \mathbf{q}(\mathbf{r}') \rangle$ for distances $\mathbf{r}-\mathbf{r}'$ along the $x$ and $y$ Cartesian directions. We average the correlation function over the whole lattice and over 10 independent cMC runs. The results are shown by circles in the top panel of Fig.~\ref{fig:xi} for $L=200$ and $\delta = 0.03$, and the colors indicate the temperature ranging from 0.409 (red) to 0.029 (blue). The black circles indicate the temperature of the peak in the specific heat for this lattice size. The lines correspond to exponential fits, from which we extract the correlation length $\xi_{qq}$ whose temperature dependence is shown in the inset. It becomes evident that there is a sudden growth of the correlation length close to the critical temperature (indicated by the dashed black line), which can be interpreted as the start of a divergence. The expected divergence in the correlation length for the Kosterlitz-Thouless transition corresponds to a power-law decay in the entire temperature region $T<T_\text{KT}$. However, in our case, a clean divergence $\xi_{qq}\rightarrow\infty$ cannot be detected due to the finite size of our system. The pink curve in the inset of Fig.~\ref{fig:xi} shows a fit of $\xi_{qq}$ in the range from $T=0.204$ to $0.084$ above the critical temperature to the aforementioned functional dependence of $\xi_{qq}(T)$ according to the Kosterlitz-Thouless theory. Apart from the high-temperature regime and the temperature region near the observed transition, the fit shows good agreement with our data. Finally, the bottom panel of Fig.~\ref{fig:xi} displays the behavior of $\xi_{qq}$ across the whole phase diagram, showing the sudden increase of $\xi_{qq}$ at the phase transition in a wider $\delta$ region.

Overall, our numerical results are consistent with a Kosterlitz-Thouless transition associated with momentum vortices, but cannot ultimately resolve the nature of this transition. If present, such a transition would constitute a rather uncommon and previously unexplored occurrence of Kosterlitz-Thouless physics. In fact, momentum degrees of freedom are quite different from the usual microscopic conditions of a Kosterlitz-Thouless transition, which, at first sight, rather speaks against a Kosterlitz-Thouless transition from momentum vortices. First, the system does not have an exact global U(1) rotation symmetry of the momentum $\mathbf{q}$. Rather, an exact energy-conserving U(1) transformation of spiral momenta is only possible in the ground state manifold. Second, from $\mathbf{q}(\mathbf{r})=\nabla \phi(\mathbf{q})$ it follows that the momentum field $\mathbf{q}(\mathbf{r})$ is curl-free, $\nabla\times \mathbf{q}(\mathbf{r})=0$, a condition that does typically not exist for more standard planar degrees of freedom such as XY spins. Third, as a result of the last property, it was shown in Ref.~\cite{Yan22} that momentum antivortices have a larger excitation energy than momentum vortices giving rise to a peculiar disparity between both vortex types. These arguments led the authors of Ref.~\cite{Yan22} to be reluctant to the possibility of a Kosterlitz-Thouless transition from momentum vortices. If our present numerical indications in favor of such a transition are true, this would indicate a striking robustness of Kosterlitz-Thouless physics. However, to eventually resolve this question, further studies are necessary.

\section{Effective rank-2 electrodynamics and pinch-points}
\label{sec:pinchpoints}

An effective continuum theory for a spiral spin liquid was recently derived in the small $\delta$ limit where the spiral ring is approximately circular~\cite{Yan22}. One of the assumptions of this theory is that the spin texture $\phi(\mathbf{r})$ is a smooth function in real space so that the momentum field $\mathbf{q}(\mathbf{r}) = \nabla \phi(\mathbf{r})$ is curl-free, $\nabla\times \mathbf{q}(\mathbf{r})=0$. Note that this excludes the possibility of spin vortices which represent a local source of curl, $\nabla\times \mathbf{q}(\mathbf{r})=\delta(\mathbf{r})$. The Hamiltonian of the continuum theory for small $\delta$ and $q=|\mathbf{q}|$ reads as
\begin{equation}\label{eq:continuum}
    \mathcal{H} = \int d^2\mathbf{r} \left( \frac{q^4}{16}-2\delta q^2\right) +  \int d^2\mathbf{r} \left( \mathcal{Q}_{\mu\nu} C_{\mu \nu \rho \sigma} \mathcal{Q}_{\rho \sigma}\right),
\end{equation}
where $C_{\mu\nu\rho\sigma} = \frac{1}{16}\left(\delta_{\mu\rho} \delta_{\nu\sigma}+\delta_{\mu \sigma} \delta_{\nu \rho}-\delta_{\mu \nu} \delta_{\rho \sigma}\right)$ is a combination of Kronecker deltas and $\mathcal{Q}_{\mu \nu} = \partial_\mu \partial_\nu \phi$ is the Hessian matrix of $\phi(\mathbf{r})$~\cite{Yan22}. The Hamiltonian in Eq.~(\ref{eq:continuum}) has two terms~\cite{Yan22}; the first term is an effective potential for spin spirals that governs the energy cost of a homogeneous spiral with momentum $\mathbf{q}$. The second term describes a spiral stiffness that captures the energy cost of deforming spiral configurations in real space. Importantly, within this effective theory a matrix-valued field $E_{\mu\nu}$ can be defined via
\begin{equation}\label{eq:electric_field}
E_{\mu\nu}=\epsilon_{\mu\rho}\epsilon_{\nu\sigma}\mathcal{Q}_{\rho \sigma},
\end{equation}
that is subject to the constraint
\begin{equation}\label{eq:gauss_law}
\partial_\mu\partial_\nu E_{\mu\nu}=0,
\end{equation}
again under the assumption that spin vortices are absent. In Eq.~(\ref{eq:electric_field}), $\epsilon_{\mu\rho}$ denotes the Levi-Civita symbol. The condition in Eq.~(\ref{eq:gauss_law}) can be interpreted as a generalized Gauss law in a charge-free rank-2 electrodynamics theory, where the emergent electric field $E_{\mu\nu}$ is a rank-2 tensor as opposed to the vector-valued electric field in conventional electrodynamics. Inspired by the unusual kinetic properties of the associated charged matter fields (in this case called {\it fractons}) which retain their mobility only within subdimensional manifolds~\cite{PhysRevB.98.125105, Pretko17, Pretko17b}, such higher-rank versions of electrodynamics have become a topical research field. For further details about the mapping between a spiral spin liquid and a rank-2 electrodynamics theory, we refer the interested reader to Ref.~\cite{Yan22}.

A characteristic feature of a rank-2 electrodynamics theory is a four-fold pinch point (4FPP) in the electric field correlation function defined by
\begin{equation}
\mathcal{C}_{EE}(\mathbf{q}) = \frac{1}{N} \sum_{\mathbf{r},\mathbf{r'}} \langle E_{xx}(\mathbf{r}) E_{yy}(\mathbf{r'}) \rangle \ e^{i \mathbf{q}\cdot (\mathbf{r'}-\mathbf{r})}.
\label{eq:efcorr}
\end{equation}
For an exact fulfillment of Eq.~(\ref{eq:gauss_law}), the 4FPP in the electric field correlator has the form
\begin{equation}
\mathcal{C}_{EE}(\mathbf{q}) \propto \frac{q_x^2 q_y^2}{|\mathbf{q}|^4}\propto \sin^2(2\theta)
\label{eq:pinch}
\end{equation}
at small $q$. In the right-most expression of Eq.~(\ref{eq:pinch}) we have used a polar representation of the momentum, i.e., $q_x = q \cos(\theta)$ and $q_y = q \sin(\theta)$. This expression makes it obvious that an ideal 4FFP does not have a dependence on $q$. Importantly, the sharpness of the non-analyticity of Eq.~(\ref{eq:pinch}) at $\mathbf{q}=0$ serves as a useful measure for the fulfillment of the generalized Gauss law in Eq.~(\ref{eq:gauss_law}). In Ref.~\cite{Yan22}, well-defined 4FPP in $\mathcal{C}_{EE}(\mathbf{q})$ have already been observed in the spiral spin liquid phase at $\delta=0.03$ using numerically-obtained spin configurations.

The mapping between a spiral spin liquid and a rank-2 electrodynamics theory, however, depends on various assumptions such as small $\delta\ll 1$ and the absence of spin vortices, both of which are not fulfilled in large parts of the phase diagram. Furthermore, the exact 4FPP shape of the electric field correlator in Eq.~(\ref{eq:pinch}) does not take into account possible thermalization problems, which, however, are known to occur in our spiral model and which limit the configuration space the system can explore. Therefore, it is {\it a priori} unclear to what extent the analogy between the spiral spin liquid and the rank-2 electrodynamics theory is valid across the phase diagram. To answer this question, we calculate the correlation function $\mathcal{C}_{EE}(\mathbf{q})$ in Eq.~(\ref{eq:efcorr}) from spin configurations obtained by cMC in the entire $\delta$-$T$ phase diagram and investigate the intactness of 4FPPs.

\begin{figure}[!t]
\centering
\includegraphics*[width=0.45\textwidth]{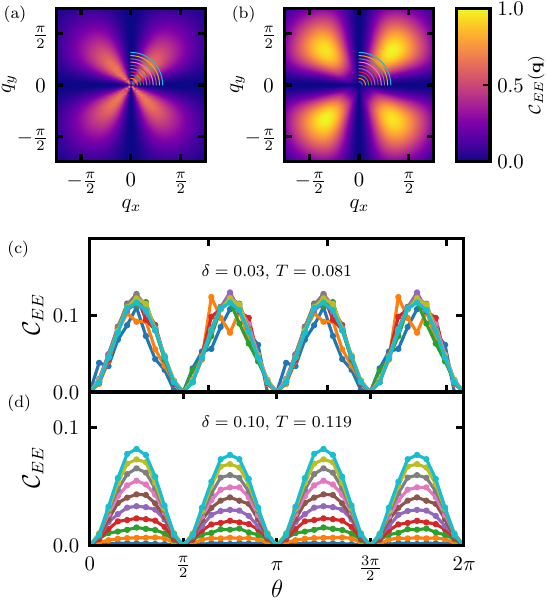}
\caption{Electric field correlator $\mathcal{C}_{EE}(\mathbf{q})$ [see Eq.~(\ref{eq:efcorr})]: (a) for $\delta=0.03$, $T=0.081$ and (b) for $\delta = 0.10$, $T=0.119$. (c) and (d) show the corresponding signal along circular paths $\theta\in[0,2\pi]$ with radii $q$ ranging from 0.1 to 1.0 at 0.1 intervals [indicated by color lines in the first quadrants of (a) and (b)].}
\label{fig:pinch}
\end{figure}

Note that even for an exact lattice realization of a rank-2 electrodynamics theory Eq.~(\ref{eq:pinch}) is only valid at small $q$ while near the boundary of the Brillouin zone, lattice effects play a role that leads to deviations from the exact 4FPP shape. For a large part of the phase diagram, we observe four symmetric lobes in $\mathcal{C}_{EE}(\mathbf{q})$ resembling the shape of 4FPPs at {\it intermediate} values of $q$. The decisive property, however, is the behavior of these lobes at $q\rightarrow0$.

We observe distinctly different behaviors in this limit across the $\delta$-$T$ phase diagram, for which we show two representative examples in  Fig.~\ref{fig:pinch}. Panels (a) and (c) display $\mathcal{C}_{EE}(\mathbf{q})$ from cMC results at $\delta = 0.03$ and $T=0.081$ (close to the phase transition between the pancake and spiral spin liquids). Ideal pinch-point singularities have an intensity that only depends on the angle $\theta$ in momentum space and not on the distance $q$ from the origin. To check this, we plot in Fig.~\ref{fig:pinch} (c) the signal along the circular paths $\theta\in[0,2\pi]$ with constant $q$ as indicated in Fig.~\ref{fig:pinch} (a). We observe that all curves overlap for different radii $q$ down to $q=0.1$, very close to the origin, indicating an intact 4FPP. On the other hand, we show in Fig.~\ref{fig:pinch} (b) and (d) results for $\delta = 0.10$ and $T=0.119$ (close to the phase transition between the paramagnet and the nematic spiral). In this case, we see that the intensity vanishes as the radius $q$ decreases, signaling the absence of a singularity. 
Such a smearing of pinch points may indicate that emergent charge fluctuations are soft and become thermally activated~\cite{Fennell09}. 

\begin{figure}[!t]
\centering
\includegraphics*[width=0.45\textwidth]{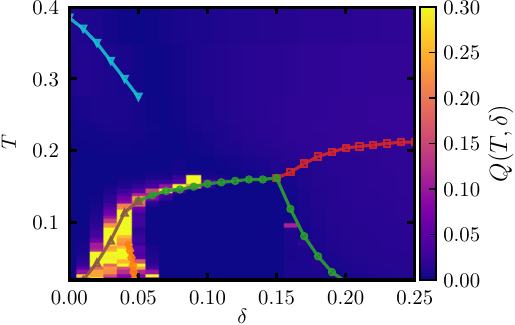}
\caption{Pinch-point singularity quality $Q$ as a function of $T$ and $\delta$ (see main text). Lines and symbols correspond to the phase transitions indicated by the heat capacity.}
\label{fig:pinch2}
\end{figure}

To investigate the behavior of 4FPP at small momenta more systematically and over the whole phase diagram, we define a quality measure of pinch points in the following way. For the 10 different values of the radii $q_i$ ranging from 0.1 to 1.0 in steps of 0.1, we fit the data to $A(q_i)\sin^2(2\theta)$ and extract $A(q_i)$. We then define our quality measure as $Q = b^2 / |a|$ where $a$ and $b$ are obtained from a fit to a linear function $A(q)=aq+b$. The choice of defining $Q$ with $b^2$ in the numerator as opposed to just $b$ is to suppress the influence of a small fluctuating background signal for $a, b \sim 0$. Then, $Q$ takes large values when $A(q)$ is approximately constant ($a\to 0$), signaling a sharp singularity. On the other hand, $Q$ takes negligible values if the slope $a$ is finite and/or $b$ vanishes (featureless electric field correlator), indicating deviations from the exact pinch point shape in Eq.~(\ref{eq:pinch}). The results for $Q$ as a function of $\delta$ and $T$ are shown in Fig.~\ref{fig:pinch2}, where it becomes clear that the quality of 4FPPs is highest around and below the phase transition into the spiral spin liquid (brown line). This demonstrates that the fluctuations in the spiral spin liquid can indeed be described by an effective Gauss law associated with a higher-rank generalization of electromagnetism. In the nematic spiral phase (below the green line), four-fold symmetric lobes with $\mathcal{C}_{EE}(\mathbf{q})\sim A\sin^2(2\theta)$ are also observed, as shown in Fig.~\ref{fig:pinch}. However, in this case, the magnitude $A=A(q)$ depends on the distance $q$ from the origin.

\section{Conclusions}
\label{sec:conc}

We have used classical Monte Carlo and molecular dynamics simulations to study the finite-temperature phase diagram of a prototypical spiral spin model on the square lattice with XY spins. We have identified a variety of interesting phases and emergent phenomena in this model. First investigating the regime of large $\delta$, at $\delta=0.25$ we found that the ground-state manifold contains spiral stripes with a wavelength of four sites, as well as colinear stripes with two-up-two-down magnetic unit cells, called the $\langle 2\rangle$ order. At finite temperatures, entropic effects select ordering wave vectors along the $x$ or $y$ directions corresponding to a discrete $\mathds{Z}_2$ symmetry breaking of the $C_4$ lattice rotation symmetry down to $C_2$. On top of this, entropy selects the colinear $\langle 2\rangle$ states over the homogeneous spin spirals which constitutes an additional $\mathds{Z}_2$ symmetry breaking related to lattice translation. We found that the two $\mathds{Z}_2$ symmetries are broken at the same temperature due to order-by-disorder effects, leading to a $\mathds{Z}_2 \times \mathds{Z}_2 = \mathds{Z}_4$ symmetry breaking into the $\langle 2\rangle$ phase. We verified that the critical exponents agree with a second-order Ashkin-Teller (or four-state Potts) phase transition. 

We found that the phase transition into the $\langle 2 \rangle$ phase persists for $0.16 \leq \delta \leq 0.25$, even though the $\langle 2 \rangle$ order is no longer part of the ground-state manifold. As a consequence, a second phase transition into a nematic spiral state arises at lower temperatures. This is an example of a re-entrance phenomenon since it implies a restoration of the broken lattice translation symmetry (while lattice rotation symmetry remains broken). We verified that this second phase transition belongs to the Ising universality class. For $0.05 < \delta \leq 0.15$, the system shows only one finite-temperature phase transition into the nematic spiral state which breaks lattice rotation symmetry. This transition is again associated with a broken $\mathds{Z}_2$ symmetry and the transition belongs to the Ising universality class. 

For $\delta < 0.05$, the peak in the specific heat splits into two peaks. A wide shoulder shifts to higher temperatures as $\delta$ decreases, while a sharp low-temperature peak decreases in intensity and shifts towards $T\rightarrow0$ as $\delta\rightarrow0$. The high-temperature feature shows an off-critical behavior, indicating a crossover into a pancake spin liquid, where spirals with all wave vectors inside the spiral ring coexist. On the other hand, the low-temperature transition leads into a spiral spin liquid phase and shows a logarithmic divergence of the specific heat in system size, consistent with a critical exponent $\alpha = 0$ and reminiscent of an Ising transition. However, no spontaneous symmetry breaking is observed across this transition and the critical exponent $\nu$ is inconsistent with an Ising transition, excluding a standard second-order transition. Instead, we found indications that the transition is driven by momentum vortices. While the density of momentum vortices shows a sudden decrease at the transition, the momentum-momentum correlation length increases sharply, pointing to the possibility of an unusual and hitherto unexplored Kosterlitz-Thouless transition from momentum degrees of freedom.

We also performed molecular dynamics calculations to characterize the dynamical spin structure factor of the different phases. The colinear $\langle 2\rangle$ phase shows low-energy spin-wave bands at all momenta on the spiral ring, contrary to the nematic spiral phase where all modes are gapped, except for the Goldstone mode. On the other hand, the pancake liquid shows abundant low-energy excitations for all wave vectors inside the spiral ring with a continuum of excitations extending to higher energies. For the spiral spin liquid, low-energy modes are only observed in the vicinity of the spiral ring while at higher energies a fading of the continuum and spectral weight mostly concentrated along well-defined spin wave bands is found.

Finally, we verified the claim of Ref.~\cite{Yan22} that the fluctuations in the spiral spin liquid can be captured by an effective rank-2 Gauss law for an emergent matrix-valued electric field as it appears in a so-called rank-2 U(1) electrodynamics theory. We took the sharpness of the characteristic 4FPP singularities in the electric-field correlator as a measure to determine the regions in the phase diagram where the mapping onto a rank-2 electrodynamics theory holds. We, indeed, verified that the sharpest 4FPP are observed in the spiral spin liquid regime, demonstrating the close connection between this phase and an emergent rank-2 electrodynamics.

In total, our work presents a coherent picture of a spin model that, even on the classical level, displays a wealth of fascinating features ranging from spiral spin liquids, re-entrance phenomena, and unusual Kosterlitz-Thouless transitions to emergent higher-rank gauge theories. While our work resolves many unanswered questions from previous studies, it also points to several aspects that deserve further investigation. For example, developing a deeper understanding and verification of the proposed Kosterlitz-Thouless transition from momentum degrees of freedom would be a worthwhile future research effort. Furthermore, the inclusion of quantum fluctuations and their effects on the rank-2 electrodynamics theory may give rise to even more fascinating emergent phenomena in this spin system.

\section*{Acknowledgements}

The authors would like to thank Johannes Knolle, Piet Brouwer, Achim Rosch,  Matthias Vojta, Randall Samien and Leo Radzihovsky for inspiring discussions. M.G.G. and A.F. would like to thank the CURTA cluster of ZEDAT, Freie Universit\"at Berlin, and the JUWELS cluster at the Forschungszentrum Jülich for computing time.

\appendix

\section{Exact vs. cMC ground-state energy}
\label{app:gse}

\begin{figure}[!t]
\centering
\includegraphics*[width=0.45\textwidth]{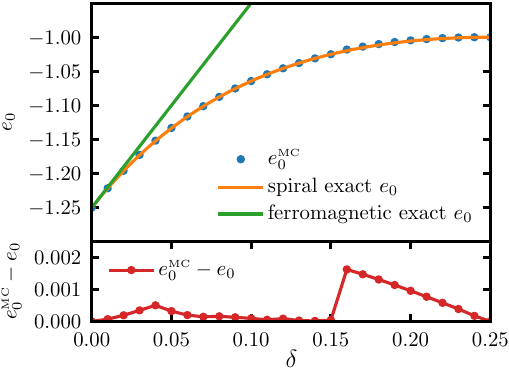}
\caption{Top panel: Ground-state energy per site $e_0$ obtained from cMC (blue circles) compared to the exact energies for the spiral ground state (orange line) and the ferromagnetic state (green line). Bottom panel: Energy difference $e_0^{\text{MC}}-e_0$ between the cMC result and the exact value.}
\label{fig:e0}
\end{figure}

The ground-state energy can be obtained via cMC as a continuation of the cooldown process to $T=0$. In Fig.~\ref{fig:e0} we compare these results for $L=200$ and periodical boundary conditions with the exact values of the ground-state energy for the spiral solutions. At low values of $\delta$, we can see that the energy of the ferromagnetic state (green line) lies very close above the spiral state. This can explain why at finite temperatures, when these ferromagnetic states can be realized, a pancake liquid emerges that displays strong ferromagnetic and spiral correlations up to the edge of the spiral ring in reciprocal space. 

On the other hand, due to the frustrating boundary conditions for incommensurate spiral phases, it is expected that the exact ground-state energy differs from the cMC result (bottom panel). However, we can see that the difference is small. Two parts show larger differences: One is the region where the system passes through the $\langle 2 \rangle$ phase. As explained in the main text, there are reminiscences of the passage through this phase in the form of small but finite $\mathcal{O}_1$ order parameter values [see Eq.~(\ref{eq:o1})] that lead to a higher ground-state energy in cMC. On the other hand, also when the system goes through the vortex network, the cMC ground-state energy differs from the exact one, showing that the dynamics of these structures are slow and that the systems need more space to accommodate properly. 

\section{Finite-size scaling and universality classes of phase transitions}
\label{app:fss}

To determine the order of the different phase transitions and the corresponding universality classes, we perform cMC calculations on a large range of lattice sizes starting from $L=4$ up to $L=80$. In some cases, the peaks in the specific heat $c_v(T)$ get sharp (thin and high) very fast with increasing system size. In such cases, it becomes difficult to obtain an accurate value of the peak, $c_v^\text{max}$, while $T_c$ is well-defined.

The correlation length of the system near the critical point diverges like $\xi \propto |T-T_c|^{-\nu}$, where $\nu$ is the critical exponent. However, in finite systems, the correlation length cannot be larger than $\sim L$. Therefore, we can assume that $|T_c(\infty) - T_c(L)|^{-\nu} \propto L$ where $T_c(L)$ is the position of the peak in the specific heat for a given linear size $L$ and $T_c(\infty)$ is the value in the thermodynamic limit $L\to \infty$. Then we obtain the finite-size scaling law~\cite{Ferdinand69, Landa76, Privman90}
\begin{equation}
T_c(L) = T_c(\infty) + a\ L^{-1/\nu}.
\label{eq:fss1}
\end{equation}
Using this, we can fit the positions of the peak in $c_v(T)$ as a function of $1/L$ and obtain $T_c(\infty)$, $a$, and $\nu$ from the fit. While typically $T_c(\infty)$ is well defined by these fits, $\nu$ tends to present larger uncertainties because the curvature of $T_c(L)$ is affected by the values on small lattices, where the effect of the boundary conditions on the state of the systems is more visible. On the other hand, the specific heat behaves like $c_v \propto |T-T_c|^{-\alpha}$ around the critical point, such that it is straightforward to see that~\cite{Privman90}
\begin{equation}
c_v^{\text{max}}(L) \propto L^{\alpha/\nu}.
\label{eq:fss2}
\end{equation}
In this case, a log-log plot allows the extraction of the quotient $\alpha/\nu$ as the slope of a linear fit, whereas in the case of Eq.~(\ref{eq:fss1}) an unknown constant $T_c(\infty)$ needs to be removed before using the same method. For this reason, the determination of $\alpha / \nu$ is more precise than that of $\nu$ from $T_c(L)$. However, obtaining a precise value for $c_v^{\text{max}}(L)$ is complicated for large lattices because the peak becomes too thin and high, leading to large differences in $c_v^{\text{max}}(L)$ for small deviations in $T_c(L)$. 

In practice, there are two different routes for calculating the specific heat and obtaining $c_v^{\text{max}}(L)$ and $T_c(L)$. One consists of calculating $c_v(T) = N(\langle e(T)^2 \rangle - \langle e(T) \rangle^2)/T^2$, where $e(T)$ is the internal energy per site and $\langle . \rangle$ is the cMC average over different Monte Carlo steps at a given temperature. Then, $c_v(T)$ is averaged over several independent runs. The second route consists of taking the discrete derivative of $e(T)$ in the temperature grid available. This is done after averaging $e(T)$ over independent runs. Ideally, both approaches should lead to the same result, but in practice, we find that the second option leads to less noisy results. Still, the specific heat calculations are not always smooth for small temperature steps $\Delta T$ because small uncertainties in the energy $e$ lead to large ones when calculating $\Delta e/\Delta T$. To overcome this, we use a Savitzky–Golay filter to smooth the curves and improve the finite-size scaling~\cite{Savitzky64}.

\begin{figure}[!t]
\centering
\includegraphics*[width=0.45\textwidth]{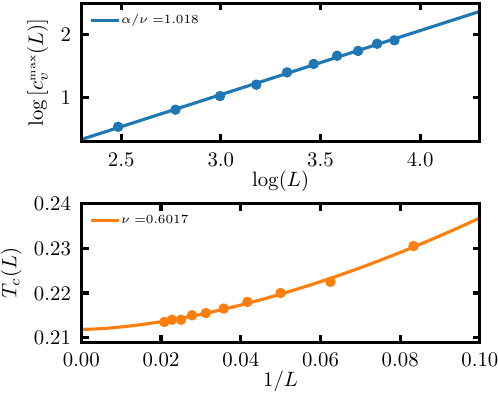}
\caption{Finite-size scaling of $c_v^{\text{max}}$ and $T_c$ for the phase transition at $\delta = 0.25$ between the paramagnetic phase and the $\langle 2\rangle$-ordered phase. Points correspond to the cMC calculations, while fits to extract the critical exponents are shown with lines (see main text).}
\label{fig:fss1}
\end{figure}

In Fig.~\ref{fig:fss1}, we show the results for the high-temperature phase transition at $\delta = 0.25$, where the system goes from a paramagnetic phase to a phase with $\langle 2 \rangle$ order. We obtain $\alpha/\nu \approx 1.0$, which is consistent with a 4-state Potts (or Ashkin-Teller) transition in two dimensions (top panel). These four possible states are realized by combinations of horizontal or vertical stripe configurations and the two possibilities for the two patterns of stripes ($\phi \phi \overline{\phi} \overline{\phi}$ and $\phi \overline{\phi} \overline{\phi} \phi$) related by a lattice translation of one nearest neighbor distance. In the bottom panel of Fig.~\ref{fig:fss1}, we see that $\nu$ takes values close to the expected $\nu = 2/3$ for the 4-state Potts transition.

\begin{figure}[!t]
\centering
\includegraphics*[width=0.45\textwidth]{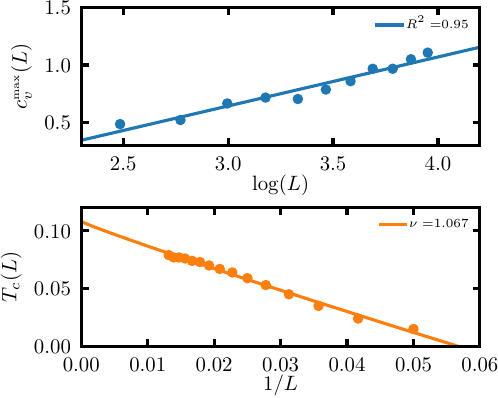}
\caption{Finite-size scaling of $c_v^{\text{max}}$ and $T_c$ for the phase transition at $\delta = 0.17$ between the $\langle 2\rangle$ phase and the nematic spiral phase. Note that the top panel uses a logarithmic scale only on the $x$-axis but not on the $y$-axis because $\alpha = 0$ for an Ising transition and the leading term becomes $c_v^{\text{max}}(L) \propto \log(L)$. The parameter that indicates the goodness of the fit, $R^2$, is shown in the legend and is close to the optimal value of 1.}
\label{fig:fss2}
\end{figure}

In Fig.~\ref{fig:fss2}, we show the results for the low-temperature phase transition between the $\langle 2 \rangle$ phase and the nematic spiral phase at $\delta = 0.17$. In this case, the peaks in the specific heat become very thin as the system size increases, and it becomes difficult to determine the height accurately (top panel). However, the critical temperature shown in the bottom panel is very well fitted by $\nu \approx 1.1$. This is consistent with $\nu = 1$ expected for an Ising transition in two dimensions. We can think of this transition coming from the nematic spiral phase at lower temperatures. Then, when the temperature increases and the system evolves to the $\langle 2 \rangle$ phase, there are two possible choices of double stripes ($\phi \phi \overline{\phi} \overline{\phi}$ and $\phi \overline{\phi} \overline{\phi} \phi$). An Ising transition is characterized by $\alpha = 0$ and the next relevant term in Eq.~(\ref{eq:fss2}) is $\propto\log(L)$. The top panel of Fig.~\ref{fig:fss2} shows that the numerical result for $c_v^{\text{max}}(L)$ as a function of $\log(L)$ is indeed well fitted by a linear function.

\begin{figure}[!t]
\centering
\includegraphics*[width=0.45\textwidth]{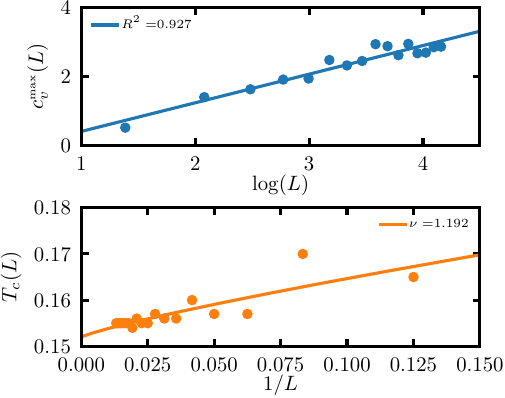}
\caption{Finite-size scaling of $c_v^{\text{max}}$ and $T_c$ for $\delta = 0.10$, extracted from the transition between the disordered and the nematic spiral state. Note that the top panel does not show a log-log plot but a lin-log plot to obtain the leading term $c_v^{\text{max}}(L) \propto \log(L)$ for an Ising transition where $\alpha=0$.}
\label{fig:fss3}
\end{figure}

In Fig.~\ref{fig:fss3} we show the results for $\delta = 0.10$, where there is a transition between a disordered and a nematic spiral state. As in the previous case, the transition is expected to be of Ising type because of the selection of stripes along the $x$ or $y$ directions. This is confirmed by the linear growth of the peak in the specific heat $c_v^{\text{max}}(L)$ as a function of $\log(L)$ (top panel). On the other hand, $T_c(L)$ shows an approximate linear behavior as a function of $1/L$, indicating $\nu = 1$, as expected for an Ising transition. The large deviation in the values $T_c(L)$ for small systems can be associated with the frustration induced by the boundary conditions, which is stronger for incommensurate spirals in the case of small lattices.

\begin{figure}[!t]
\centering
\includegraphics*[width=0.45\textwidth]{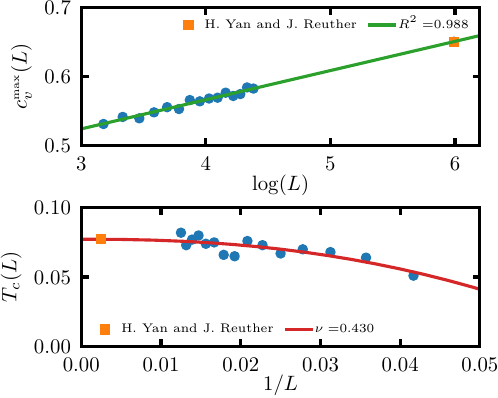}
\caption{Finite-size scaling of $c_v^{\text{max}}$ and $T_c$ for the low-temperature peak at $\delta = 0.03$, where the system evolves from the pancake to the spiral spin liquid. The orange square corresponds to the result of Ref.~\cite{Yan22} for $L=400$.}
\label{fig:fss4}
\end{figure}

Finally, in Fig.~\ref{fig:fss4} we show the finite-size scaling for the low-temperature peak at $\delta = 0.03$, where the system goes from the pancake liquid to the spiral spin liquid. In this case, we observe that $c_v^{\text{max}}(L)$ diverges logarithmically with system size, indicating that the corresponding critical exponent vanishes, $\alpha=0$. However, $T_c(L)$ does not evolve linearly with $1/L$, as expected for an Ising transition ($\alpha = 0$). Instead, the critical exponent is closer to $\nu=1/2$, which corresponds to a linear behavior in $L^{-1/\nu}=N^{-1}$. As explained in the main text, this transition is connected to the proliferation of momentum vortices and antivortices and may be a Kosterlitz-Thouless transition of momentum vortices. However, a more thorough investigation is needed to confirm this scenario.

\section{Transition between the spiral spin liquid and nematic spiral phase}
\label{app:zoomin}

\begin{figure}[!t]
\centering
\includegraphics*[width=0.45\textwidth]{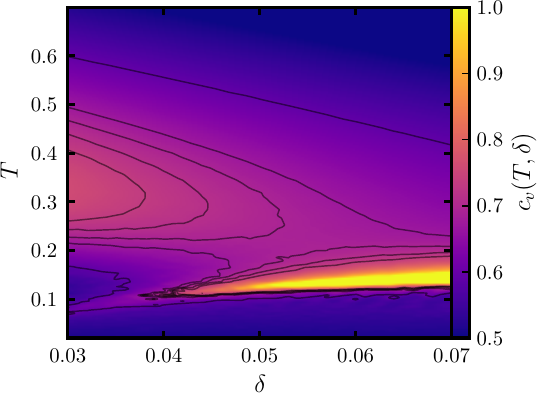}
\caption{Specific heat $c_v$ as a function of $\delta$ and $T$ in a smaller region of the $\delta$-$T$ space than in Fig.~\ref{fig:phased}, for lattice size $L=200$ and periodical boundary conditions.}
\label{fig:zoomin}
\end{figure}

As discussed in the previous section and in the main text, the phase transition for $\delta>0.05$ belongs to the Ising universality class and leads to a nematic stripe spiral order. On the other hand, for $\delta < 0.05$ the transition does not fit in any universality class and presents no symmetry breaking. Since the two low-temperature phases are fundamentally different in terms of broken symmetries, a phase transition has to exist between them. However, as shown in Fig.~\ref{fig:phased} and in Fig.~\ref{fig:zoomin}, such a phase transition is not observed in the specific heat. For the enlarged view in Fig.~\ref{fig:zoomin}, we performed independent cMC runs for $\delta$ between 0.03 and 0.07 in 0.001 steps. At $\delta = 0.07$, there is a clear transition from a disordered state into a nematic spiral state. As $\delta$ is lowered, the transition is weakened, i.e., the peak decreases and shifts to lower temperatures. Furthermore, a wide shoulder emerges at higher temperatures (as evidenced by the contour lines). The latter indicates the onset of correlations leading to the pancake liquid phase. The low-energy transition indicates the passage into the spiral spin liquid without symmetry breaking. As shown in the previous section, all along the low-temperature phase transition, the specific heat diverges logarithmically. The fact that the specific heat calculations cannot resolve this transition even with this small $\delta$ step size indicates that the single-spin update algorithm might not be sufficient to capture it and further studies are needed to confirm its presence. Nonetheless, as discussed in the main text, the transition can be detected via the calculation of appropriate order parameters and evidenced by energy-level crossings in $\delta$-sweeps.

\bibliography{papers}

\begin{thebibliography}{66}%
\makeatletter
\providecommand \@ifxundefined [1]{%
 \@ifx{#1\undefined}
}%
\providecommand \@ifnum [1]{%
 \ifnum #1\expandafter \@firstoftwo
 \else \expandafter \@secondoftwo
 \fi
}%
\providecommand \@ifx [1]{%
 \ifx #1\expandafter \@firstoftwo
 \else \expandafter \@secondoftwo
 \fi
}%
\providecommand \natexlab [1]{#1}%
\providecommand \enquote  [1]{``#1''}%
\providecommand \bibnamefont  [1]{#1}%
\providecommand \bibfnamefont [1]{#1}%
\providecommand \citenamefont [1]{#1}%
\providecommand \href@noop [0]{\@secondoftwo}%
\providecommand \href [0]{\begingroup \@sanitize@url \@href}%
\providecommand \@href[1]{\@@startlink{#1}\@@href}%
\providecommand \@@href[1]{\endgroup#1\@@endlink}%
\providecommand \@sanitize@url [0]{\catcode `\\12\catcode `\$12\catcode `\&12\catcode `\#12\catcode `\^12\catcode `\_12\catcode `\%12\relax}%
\providecommand \@@startlink[1]{}%
\providecommand \@@endlink[0]{}%
\providecommand \url  [0]{\begingroup\@sanitize@url \@url }%
\providecommand \@url [1]{\endgroup\@href {#1}{\urlprefix }}%
\providecommand \urlprefix  [0]{URL }%
\providecommand \Eprint [0]{\href }%
\providecommand \doibase [0]{https://doi.org/}%
\providecommand \selectlanguage [0]{\@gobble}%
\providecommand \bibinfo  [0]{\@secondoftwo}%
\providecommand \bibfield  [0]{\@secondoftwo}%
\providecommand \translation [1]{[#1]}%
\providecommand \BibitemOpen [0]{}%
\providecommand \bibitemStop [0]{}%
\providecommand \bibitemNoStop [0]{.\EOS\space}%
\providecommand \EOS [0]{\spacefactor3000\relax}%
\providecommand \BibitemShut  [1]{\csname bibitem#1\endcsname}%
\let\auto@bib@innerbib\@empty
\bibitem [{\citenamefont {Balents}(2010)}]{Balents10}%
  \BibitemOpen
  \bibfield  {author} {\bibinfo {author} {\bibfnamefont {L.}~\bibnamefont {Balents}},\ }\bibfield  {title} {\bibinfo {title} {{Spin liquids in frustrated magnets}},\ }\href {https://doi.org/10.1038/nature08917} {\bibfield  {journal} {\bibinfo  {journal} {Nature}\ }\textbf {\bibinfo {volume} {464}},\ \bibinfo {pages} {199} (\bibinfo {year} {2010})}\BibitemShut {NoStop}%
\bibitem [{\citenamefont {Savary}\ and\ \citenamefont {Balents}(2016)}]{Savary17}%
  \BibitemOpen
  \bibfield  {author} {\bibinfo {author} {\bibfnamefont {L.}~\bibnamefont {Savary}}\ and\ \bibinfo {author} {\bibfnamefont {L.}~\bibnamefont {Balents}},\ }\bibfield  {title} {\bibinfo {title} {{Quantum spin liquids: a review}},\ }\href {https://doi.org/10.1088/0034-4885/80/1/016502} {\bibfield  {journal} {\bibinfo  {journal} {Reports on Progress in Physics}\ }\textbf {\bibinfo {volume} {80}},\ \bibinfo {pages} {016502} (\bibinfo {year} {2016})}\BibitemShut {NoStop}%
\bibitem [{\citenamefont {Broholm}\ \emph {et~al.}(2020)\citenamefont {Broholm}, \citenamefont {Cava}, \citenamefont {Kivelson}, \citenamefont {Nocera}, \citenamefont {Norman},\ and\ \citenamefont {Senthil}}]{Broholm20}%
  \BibitemOpen
  \bibfield  {author} {\bibinfo {author} {\bibfnamefont {C.}~\bibnamefont {Broholm}}, \bibinfo {author} {\bibfnamefont {R.~J.}\ \bibnamefont {Cava}}, \bibinfo {author} {\bibfnamefont {S.~A.}\ \bibnamefont {Kivelson}}, \bibinfo {author} {\bibfnamefont {D.~G.}\ \bibnamefont {Nocera}}, \bibinfo {author} {\bibfnamefont {M.~R.}\ \bibnamefont {Norman}},\ and\ \bibinfo {author} {\bibfnamefont {T.}~\bibnamefont {Senthil}},\ }\bibfield  {title} {\bibinfo {title} {{Quantum spin liquids}},\ }\href {https://doi.org/10.1126/science.aay0668} {\bibfield  {journal} {\bibinfo  {journal} {Science}\ }\textbf {\bibinfo {volume} {367}},\ \bibinfo {pages} {eaay0668} (\bibinfo {year} {2020})}\BibitemShut {NoStop}%
\bibitem [{\citenamefont {Bramwell}\ and\ \citenamefont {Gingras}(2001)}]{Bramwell01}%
  \BibitemOpen
  \bibfield  {author} {\bibinfo {author} {\bibfnamefont {S.~T.}\ \bibnamefont {Bramwell}}\ and\ \bibinfo {author} {\bibfnamefont {M.~J.~P.}\ \bibnamefont {Gingras}},\ }\bibfield  {title} {\bibinfo {title} {{Spin Ice State in Frustrated Magnetic Pyrochlore Materials}},\ }\href {https://doi.org/10.1126/science.1064761} {\bibfield  {journal} {\bibinfo  {journal} {Science}\ }\textbf {\bibinfo {volume} {294}},\ \bibinfo {pages} {1495} (\bibinfo {year} {2001})}\BibitemShut {NoStop}%
\bibitem [{\citenamefont {Castelnovo}\ \emph {et~al.}(2008)\citenamefont {Castelnovo}, \citenamefont {Moessner},\ and\ \citenamefont {Sondhi}}]{Castelnovo08}%
  \BibitemOpen
  \bibfield  {author} {\bibinfo {author} {\bibfnamefont {C.}~\bibnamefont {Castelnovo}}, \bibinfo {author} {\bibfnamefont {R.}~\bibnamefont {Moessner}},\ and\ \bibinfo {author} {\bibfnamefont {S.~L.}\ \bibnamefont {Sondhi}},\ }\bibfield  {title} {\bibinfo {title} {{Magnetic monopoles in spin ice}},\ }\href {https://doi.org/10.1038/nature06433} {\bibfield  {journal} {\bibinfo  {journal} {Nature}\ }\textbf {\bibinfo {volume} {451}},\ \bibinfo {pages} {42} (\bibinfo {year} {2008})}\BibitemShut {NoStop}%
\bibitem [{\citenamefont {Fennell}\ \emph {et~al.}(2009)\citenamefont {Fennell}, \citenamefont {Deen}, \citenamefont {Wildes}, \citenamefont {Schmalzl}, \citenamefont {Prabhakaran}, \citenamefont {Boothroyd}, \citenamefont {Aldus}, \citenamefont {McMorrow},\ and\ \citenamefont {Bramwell}}]{Fennell09}%
  \BibitemOpen
  \bibfield  {author} {\bibinfo {author} {\bibfnamefont {T.}~\bibnamefont {Fennell}}, \bibinfo {author} {\bibfnamefont {P.~P.}\ \bibnamefont {Deen}}, \bibinfo {author} {\bibfnamefont {A.~R.}\ \bibnamefont {Wildes}}, \bibinfo {author} {\bibfnamefont {K.}~\bibnamefont {Schmalzl}}, \bibinfo {author} {\bibfnamefont {D.}~\bibnamefont {Prabhakaran}}, \bibinfo {author} {\bibfnamefont {A.~T.}\ \bibnamefont {Boothroyd}}, \bibinfo {author} {\bibfnamefont {R.~J.}\ \bibnamefont {Aldus}}, \bibinfo {author} {\bibfnamefont {D.~F.}\ \bibnamefont {McMorrow}},\ and\ \bibinfo {author} {\bibfnamefont {S.~T.}\ \bibnamefont {Bramwell}},\ }\bibfield  {title} {\bibinfo {title} {{Magnetic Coulomb Phase in the Spin Ice Ho$_2$Ti$_2$O$_7$}},\ }\href {https://doi.org/10.1126/science.1177582} {\bibfield  {journal} {\bibinfo  {journal} {Science}\ }\textbf {\bibinfo {volume} {326}},\ \bibinfo {pages} {415} (\bibinfo {year} {2009})}\BibitemShut {NoStop}%
\bibitem [{\citenamefont {Morris}\ \emph {et~al.}(2009)\citenamefont {Morris}, \citenamefont {Tennant}, \citenamefont {Grigera}, \citenamefont {Klemke}, \citenamefont {Castelnovo}, \citenamefont {Moessner}, \citenamefont {Czternasty}, \citenamefont {Meissner}, \citenamefont {Rule}, \citenamefont {Hoffmann}, \citenamefont {Kiefer}, \citenamefont {Gerischer}, \citenamefont {Slobinsky},\ and\ \citenamefont {Perry}}]{Morris09}%
  \BibitemOpen
  \bibfield  {author} {\bibinfo {author} {\bibfnamefont {D.~J.~P.}\ \bibnamefont {Morris}}, \bibinfo {author} {\bibfnamefont {D.~A.}\ \bibnamefont {Tennant}}, \bibinfo {author} {\bibfnamefont {S.~A.}\ \bibnamefont {Grigera}}, \bibinfo {author} {\bibfnamefont {B.}~\bibnamefont {Klemke}}, \bibinfo {author} {\bibfnamefont {C.}~\bibnamefont {Castelnovo}}, \bibinfo {author} {\bibfnamefont {R.}~\bibnamefont {Moessner}}, \bibinfo {author} {\bibfnamefont {C.}~\bibnamefont {Czternasty}}, \bibinfo {author} {\bibfnamefont {M.}~\bibnamefont {Meissner}}, \bibinfo {author} {\bibfnamefont {K.~C.}\ \bibnamefont {Rule}}, \bibinfo {author} {\bibfnamefont {J.-U.}\ \bibnamefont {Hoffmann}}, \bibinfo {author} {\bibfnamefont {K.}~\bibnamefont {Kiefer}}, \bibinfo {author} {\bibfnamefont {S.}~\bibnamefont {Gerischer}}, \bibinfo {author} {\bibfnamefont {D.}~\bibnamefont {Slobinsky}},\ and\ \bibinfo {author} {\bibfnamefont {R.~S.}\ \bibnamefont {Perry}},\ }\bibfield  {title} {\bibinfo {title} {{Dirac Strings and Magnetic Monopoles in
  the Spin Ice Dy$_2$Ti$_2$O$_7$}},\ }\href {https://doi.org/10.1126/science.1178868} {\bibfield  {journal} {\bibinfo  {journal} {Science}\ }\textbf {\bibinfo {volume} {326}},\ \bibinfo {pages} {411} (\bibinfo {year} {2009})}\BibitemShut {NoStop}%
\bibitem [{\citenamefont {Alzate-Cardona}\ \emph {et~al.}(2019)\citenamefont {Alzate-Cardona}, \citenamefont {Sabogal-Suárez}, \citenamefont {Evans},\ and\ \citenamefont {Restrepo-Parra}}]{Alzate19}%
  \BibitemOpen
  \bibfield  {author} {\bibinfo {author} {\bibfnamefont {J.~D.}\ \bibnamefont {Alzate-Cardona}}, \bibinfo {author} {\bibfnamefont {D.}~\bibnamefont {Sabogal-Suárez}}, \bibinfo {author} {\bibfnamefont {R.~F.~L.}\ \bibnamefont {Evans}},\ and\ \bibinfo {author} {\bibfnamefont {E.}~\bibnamefont {Restrepo-Parra}},\ }\bibfield  {title} {\bibinfo {title} {{Optimal phase space sampling for Monte Carlo simulations of Heisenberg spin systems}},\ }\href {https://doi.org/10.1088/1361-648X/aaf852} {\bibfield  {journal} {\bibinfo  {journal} {J. Phys. Condens. Matter}\ }\textbf {\bibinfo {volume} {31}},\ \bibinfo {pages} {095802} (\bibinfo {year} {2019})}\BibitemShut {NoStop}%
\bibitem [{\citenamefont {Chalker}\ \emph {et~al.}(1992)\citenamefont {Chalker}, \citenamefont {Holdsworth},\ and\ \citenamefont {Shender}}]{Chalker92}%
  \BibitemOpen
  \bibfield  {author} {\bibinfo {author} {\bibfnamefont {J.~T.}\ \bibnamefont {Chalker}}, \bibinfo {author} {\bibfnamefont {P.~C.~W.}\ \bibnamefont {Holdsworth}},\ and\ \bibinfo {author} {\bibfnamefont {E.~F.}\ \bibnamefont {Shender}},\ }\bibfield  {title} {\bibinfo {title} {{Hidden order in a frustrated system: Properties of the Heisenberg Kagom\'e antiferromagnet}},\ }\href {https://doi.org/10.1103/PhysRevLett.68.855} {\bibfield  {journal} {\bibinfo  {journal} {Phys. Rev. Lett.}\ }\textbf {\bibinfo {volume} {68}},\ \bibinfo {pages} {855} (\bibinfo {year} {1992})}\BibitemShut {NoStop}%
\bibitem [{\citenamefont {Huse}\ and\ \citenamefont {Rutenberg}(1992)}]{Huse92}%
  \BibitemOpen
  \bibfield  {author} {\bibinfo {author} {\bibfnamefont {D.~A.}\ \bibnamefont {Huse}}\ and\ \bibinfo {author} {\bibfnamefont {A.~D.}\ \bibnamefont {Rutenberg}},\ }\bibfield  {title} {\bibinfo {title} {{Classical antiferromagnets on the Kagom\'e lattice}},\ }\href {https://doi.org/10.1103/PhysRevB.45.7536} {\bibfield  {journal} {\bibinfo  {journal} {Phys. Rev. B}\ }\textbf {\bibinfo {volume} {45}},\ \bibinfo {pages} {7536} (\bibinfo {year} {1992})}\BibitemShut {NoStop}%
\bibitem [{\citenamefont {Isakov}\ \emph {et~al.}(2004)\citenamefont {Isakov}, \citenamefont {Gregor}, \citenamefont {Moessner},\ and\ \citenamefont {Sondhi}}]{Isakov04}%
  \BibitemOpen
  \bibfield  {author} {\bibinfo {author} {\bibfnamefont {S.~V.}\ \bibnamefont {Isakov}}, \bibinfo {author} {\bibfnamefont {K.}~\bibnamefont {Gregor}}, \bibinfo {author} {\bibfnamefont {R.}~\bibnamefont {Moessner}},\ and\ \bibinfo {author} {\bibfnamefont {S.~L.}\ \bibnamefont {Sondhi}},\ }\bibfield  {title} {\bibinfo {title} {{Dipolar Spin Correlations in Classical Pyrochlore Magnets}},\ }\href {https://doi.org/10.1103/PhysRevLett.93.167204} {\bibfield  {journal} {\bibinfo  {journal} {Phys. Rev. Lett.}\ }\textbf {\bibinfo {volume} {93}},\ \bibinfo {pages} {167204} (\bibinfo {year} {2004})}\BibitemShut {NoStop}%
\bibitem [{\citenamefont {Moessner}\ and\ \citenamefont {Chalker}(1998{\natexlab{a}})}]{Moessner98}%
  \BibitemOpen
  \bibfield  {author} {\bibinfo {author} {\bibfnamefont {R.}~\bibnamefont {Moessner}}\ and\ \bibinfo {author} {\bibfnamefont {J.~T.}\ \bibnamefont {Chalker}},\ }\bibfield  {title} {\bibinfo {title} {{Properties of a Classical Spin Liquid: The Heisenberg Pyrochlore Antiferromagnet}},\ }\href {https://doi.org/10.1103/PhysRevLett.80.2929} {\bibfield  {journal} {\bibinfo  {journal} {Phys. Rev. Lett.}\ }\textbf {\bibinfo {volume} {80}},\ \bibinfo {pages} {2929} (\bibinfo {year} {1998}{\natexlab{a}})}\BibitemShut {NoStop}%
\bibitem [{\citenamefont {Hopkinson}\ \emph {et~al.}(2007)\citenamefont {Hopkinson}, \citenamefont {Isakov}, \citenamefont {Kee},\ and\ \citenamefont {Kim}}]{Hopkinson07}%
  \BibitemOpen
  \bibfield  {author} {\bibinfo {author} {\bibfnamefont {J.~M.}\ \bibnamefont {Hopkinson}}, \bibinfo {author} {\bibfnamefont {S.~V.}\ \bibnamefont {Isakov}}, \bibinfo {author} {\bibfnamefont {H.-Y.}\ \bibnamefont {Kee}},\ and\ \bibinfo {author} {\bibfnamefont {Y.~B.}\ \bibnamefont {Kim}},\ }\bibfield  {title} {\bibinfo {title} {{Classical Antiferromagnet on a Hyperkagome Lattice}},\ }\href {https://doi.org/10.1103/PhysRevLett.99.037201} {\bibfield  {journal} {\bibinfo  {journal} {Phys. Rev. Lett.}\ }\textbf {\bibinfo {volume} {99}},\ \bibinfo {pages} {037201} (\bibinfo {year} {2007})}\BibitemShut {NoStop}%
\bibitem [{\citenamefont {Davier}\ \emph {et~al.}(2023)\citenamefont {Davier}, \citenamefont {G\'omez~Albarrac\'{\i}n}, \citenamefont {Rosales},\ and\ \citenamefont {Pujol}}]{Davier23}%
  \BibitemOpen
  \bibfield  {author} {\bibinfo {author} {\bibfnamefont {N.}~\bibnamefont {Davier}}, \bibinfo {author} {\bibfnamefont {F.~A.}\ \bibnamefont {G\'omez~Albarrac\'{\i}n}}, \bibinfo {author} {\bibfnamefont {H.~D.}\ \bibnamefont {Rosales}},\ and\ \bibinfo {author} {\bibfnamefont {P.}~\bibnamefont {Pujol}},\ }\bibfield  {title} {\bibinfo {title} {{Combined approach to analyze and classify families of classical spin liquids}},\ }\href {https://doi.org/10.1103/PhysRevB.108.054408} {\bibfield  {journal} {\bibinfo  {journal} {Phys. Rev. B}\ }\textbf {\bibinfo {volume} {108}},\ \bibinfo {pages} {054408} (\bibinfo {year} {2023})}\BibitemShut {NoStop}%
\bibitem [{\citenamefont {Yan}\ \emph {et~al.}(2023{\natexlab{a}})\citenamefont {Yan}, \citenamefont {Benton}, \citenamefont {Moessner},\ and\ \citenamefont {Nevidomskyy}}]{yan2023classification}%
  \BibitemOpen
  \bibfield  {author} {\bibinfo {author} {\bibfnamefont {H.}~\bibnamefont {Yan}}, \bibinfo {author} {\bibfnamefont {O.}~\bibnamefont {Benton}}, \bibinfo {author} {\bibfnamefont {R.}~\bibnamefont {Moessner}},\ and\ \bibinfo {author} {\bibfnamefont {A.~H.}\ \bibnamefont {Nevidomskyy}},\ }\href@noop {} {\bibinfo {title} {{Classification of Classical Spin Liquids: Typology and Resulting Landscape}}} (\bibinfo {year} {2023}{\natexlab{a}}),\ \Eprint {https://arxiv.org/abs/2305.00155} {arXiv:2305.00155 [cond-mat.str-el]} \BibitemShut {NoStop}%
\bibitem [{\citenamefont {Yan}\ \emph {et~al.}(2023{\natexlab{b}})\citenamefont {Yan}, \citenamefont {Benton}, \citenamefont {Nevidomskyy},\ and\ \citenamefont {Moessner}}]{yan2023classification2}%
  \BibitemOpen
  \bibfield  {author} {\bibinfo {author} {\bibfnamefont {H.}~\bibnamefont {Yan}}, \bibinfo {author} {\bibfnamefont {O.}~\bibnamefont {Benton}}, \bibinfo {author} {\bibfnamefont {A.~H.}\ \bibnamefont {Nevidomskyy}},\ and\ \bibinfo {author} {\bibfnamefont {R.}~\bibnamefont {Moessner}},\ }\href@noop {} {\bibinfo {title} {{Classification of Classical Spin Liquids: Detailed Formalism and Suite of Examples}}} (\bibinfo {year} {2023}{\natexlab{b}}),\ \Eprint {https://arxiv.org/abs/2305.19189} {arXiv:2305.19189 [cond-mat.str-el]} \BibitemShut {NoStop}%
\bibitem [{\citenamefont {Fang}\ \emph {et~al.}(2023)\citenamefont {Fang}, \citenamefont {Cano}, \citenamefont {Nevidomskyy},\ and\ \citenamefont {Yan}}]{fang2023classification3}%
  \BibitemOpen
  \bibfield  {author} {\bibinfo {author} {\bibfnamefont {Y.}~\bibnamefont {Fang}}, \bibinfo {author} {\bibfnamefont {J.}~\bibnamefont {Cano}}, \bibinfo {author} {\bibfnamefont {A.~H.}\ \bibnamefont {Nevidomskyy}},\ and\ \bibinfo {author} {\bibfnamefont {H.}~\bibnamefont {Yan}},\ }\href@noop {} {\bibinfo {title} {{Classification of Classical Spin Liquids: Topological Quantum Chemistry and Crystalline Symmetry}}} (\bibinfo {year} {2023}),\ \Eprint {https://arxiv.org/abs/2309.12652} {arXiv:2309.12652 [cond-mat.str-el]} \BibitemShut {NoStop}%
\bibitem [{\citenamefont {Sachdev}(1992)}]{Sachdev92}%
  \BibitemOpen
  \bibfield  {author} {\bibinfo {author} {\bibfnamefont {S.}~\bibnamefont {Sachdev}},\ }\bibfield  {title} {\bibinfo {title} {{Kagom\'e- and triangular-lattice Heisenberg antiferromagnets: Ordering from quantum fluctuations and quantum-disordered ground states with unconfined bosonic spinons}},\ }\href {https://doi.org/10.1103/PhysRevB.45.12377} {\bibfield  {journal} {\bibinfo  {journal} {Phys. Rev. B}\ }\textbf {\bibinfo {volume} {45}},\ \bibinfo {pages} {12377} (\bibinfo {year} {1992})}\BibitemShut {NoStop}%
\bibitem [{\citenamefont {Lecheminant}\ \emph {et~al.}(1997)\citenamefont {Lecheminant}, \citenamefont {Bernu}, \citenamefont {Lhuillier}, \citenamefont {Pierre},\ and\ \citenamefont {Sindzingre}}]{Lecheminant97}%
  \BibitemOpen
  \bibfield  {author} {\bibinfo {author} {\bibfnamefont {P.}~\bibnamefont {Lecheminant}}, \bibinfo {author} {\bibfnamefont {B.}~\bibnamefont {Bernu}}, \bibinfo {author} {\bibfnamefont {C.}~\bibnamefont {Lhuillier}}, \bibinfo {author} {\bibfnamefont {L.}~\bibnamefont {Pierre}},\ and\ \bibinfo {author} {\bibfnamefont {P.}~\bibnamefont {Sindzingre}},\ }\bibfield  {title} {\bibinfo {title} {{Order versus disorder in the quantum Heisenberg antiferromagnet on the kagom\'e lattice using exact spectra analysis}},\ }\href {https://doi.org/10.1103/PhysRevB.56.2521} {\bibfield  {journal} {\bibinfo  {journal} {Phys. Rev. B}\ }\textbf {\bibinfo {volume} {56}},\ \bibinfo {pages} {2521} (\bibinfo {year} {1997})}\BibitemShut {NoStop}%
\bibitem [{\citenamefont {Yan}\ \emph {et~al.}(2011)\citenamefont {Yan}, \citenamefont {Huse},\ and\ \citenamefont {White}}]{Yan11}%
  \BibitemOpen
  \bibfield  {author} {\bibinfo {author} {\bibfnamefont {S.}~\bibnamefont {Yan}}, \bibinfo {author} {\bibfnamefont {D.~A.}\ \bibnamefont {Huse}},\ and\ \bibinfo {author} {\bibfnamefont {S.~R.}\ \bibnamefont {White}},\ }\bibfield  {title} {\bibinfo {title} {{Spin-Liquid Ground State of the S = 1/2 Kagome Heisenberg Antiferromagnet}},\ }\href {https://doi.org/10.1126/science.1201080} {\bibfield  {journal} {\bibinfo  {journal} {Science}\ }\textbf {\bibinfo {volume} {332}},\ \bibinfo {pages} {1173} (\bibinfo {year} {2011})}\BibitemShut {NoStop}%
\bibitem [{\citenamefont {Messio}\ \emph {et~al.}(2012)\citenamefont {Messio}, \citenamefont {Bernu},\ and\ \citenamefont {Lhuillier}}]{Messio12}%
  \BibitemOpen
  \bibfield  {author} {\bibinfo {author} {\bibfnamefont {L.}~\bibnamefont {Messio}}, \bibinfo {author} {\bibfnamefont {B.}~\bibnamefont {Bernu}},\ and\ \bibinfo {author} {\bibfnamefont {C.}~\bibnamefont {Lhuillier}},\ }\bibfield  {title} {\bibinfo {title} {{Kagome Antiferromagnet: A Chiral Topological Spin Liquid?}},\ }\href {https://doi.org/10.1103/PhysRevLett.108.207204} {\bibfield  {journal} {\bibinfo  {journal} {Phys. Rev. Lett.}\ }\textbf {\bibinfo {volume} {108}},\ \bibinfo {pages} {207204} (\bibinfo {year} {2012})}\BibitemShut {NoStop}%
\bibitem [{\citenamefont {Gingras}\ and\ \citenamefont {McClarty}(2014)}]{Gingras14}%
  \BibitemOpen
  \bibfield  {author} {\bibinfo {author} {\bibfnamefont {M.~J.~P.}\ \bibnamefont {Gingras}}\ and\ \bibinfo {author} {\bibfnamefont {P.~A.}\ \bibnamefont {McClarty}},\ }\bibfield  {title} {\bibinfo {title} {{Quantum spin ice: a search for gapless quantum spin liquids in pyrochlore magnets}},\ }\href {https://doi.org/10.1088/0034-4885/77/5/056501} {\bibfield  {journal} {\bibinfo  {journal} {Reports on Progress in Physics}\ }\textbf {\bibinfo {volume} {77}},\ \bibinfo {pages} {056501} (\bibinfo {year} {2014})}\BibitemShut {NoStop}%
\bibitem [{\citenamefont {Huse}\ \emph {et~al.}(2003)\citenamefont {Huse}, \citenamefont {Krauth}, \citenamefont {Moessner},\ and\ \citenamefont {Sondhi}}]{Huse03}%
  \BibitemOpen
  \bibfield  {author} {\bibinfo {author} {\bibfnamefont {D.~A.}\ \bibnamefont {Huse}}, \bibinfo {author} {\bibfnamefont {W.}~\bibnamefont {Krauth}}, \bibinfo {author} {\bibfnamefont {R.}~\bibnamefont {Moessner}},\ and\ \bibinfo {author} {\bibfnamefont {S.~L.}\ \bibnamefont {Sondhi}},\ }\bibfield  {title} {\bibinfo {title} {{Coulomb and Liquid Dimer Models in Three Dimensions}},\ }\href {https://doi.org/10.1103/PhysRevLett.91.167004} {\bibfield  {journal} {\bibinfo  {journal} {Phys. Rev. Lett.}\ }\textbf {\bibinfo {volume} {91}},\ \bibinfo {pages} {167004} (\bibinfo {year} {2003})}\BibitemShut {NoStop}%
\bibitem [{\citenamefont {Hermele}\ \emph {et~al.}(2004)\citenamefont {Hermele}, \citenamefont {Fisher},\ and\ \citenamefont {Balents}}]{Hermele04}%
  \BibitemOpen
  \bibfield  {author} {\bibinfo {author} {\bibfnamefont {M.}~\bibnamefont {Hermele}}, \bibinfo {author} {\bibfnamefont {M.~P.~A.}\ \bibnamefont {Fisher}},\ and\ \bibinfo {author} {\bibfnamefont {L.}~\bibnamefont {Balents}},\ }\bibfield  {title} {\bibinfo {title} {{Pyrochlore photons: The $U(1)$ spin liquid in a $S=\frac{1}{2}$ three-dimensional frustrated magnet}},\ }\href {https://doi.org/10.1103/PhysRevB.69.064404} {\bibfield  {journal} {\bibinfo  {journal} {Phys. Rev. B}\ }\textbf {\bibinfo {volume} {69}},\ \bibinfo {pages} {064404} (\bibinfo {year} {2004})}\BibitemShut {NoStop}%
\bibitem [{\citenamefont {Bergman}\ \emph {et~al.}(2007)\citenamefont {Bergman}, \citenamefont {Alicea}, \citenamefont {Gull}, \citenamefont {Trebst},\ and\ \citenamefont {Balents}}]{Bergman07}%
  \BibitemOpen
  \bibfield  {author} {\bibinfo {author} {\bibfnamefont {D.}~\bibnamefont {Bergman}}, \bibinfo {author} {\bibfnamefont {J.}~\bibnamefont {Alicea}}, \bibinfo {author} {\bibfnamefont {E.}~\bibnamefont {Gull}}, \bibinfo {author} {\bibfnamefont {S.}~\bibnamefont {Trebst}},\ and\ \bibinfo {author} {\bibfnamefont {L.}~\bibnamefont {Balents}},\ }\bibfield  {title} {\bibinfo {title} {{Order-by-disorder and spiral spin-liquid in frustrated diamond-lattice antiferromagnets}},\ }\href {https://doi.org/10.1038/nphys622} {\bibfield  {journal} {\bibinfo  {journal} {Nature Physics}\ }\textbf {\bibinfo {volume} {3}},\ \bibinfo {pages} {487} (\bibinfo {year} {2007})}\BibitemShut {NoStop}%
\bibitem [{\citenamefont {Mulder}\ \emph {et~al.}(2010)\citenamefont {Mulder}, \citenamefont {Ganesh}, \citenamefont {Capriotti},\ and\ \citenamefont {Paramekanti}}]{Mulder10}%
  \BibitemOpen
  \bibfield  {author} {\bibinfo {author} {\bibfnamefont {A.}~\bibnamefont {Mulder}}, \bibinfo {author} {\bibfnamefont {R.}~\bibnamefont {Ganesh}}, \bibinfo {author} {\bibfnamefont {L.}~\bibnamefont {Capriotti}},\ and\ \bibinfo {author} {\bibfnamefont {A.}~\bibnamefont {Paramekanti}},\ }\bibfield  {title} {\bibinfo {title} {{Spiral order by disorder and lattice nematic order in a frustrated Heisenberg antiferromagnet on the honeycomb lattice}},\ }\href {https://doi.org/10.1103/PhysRevB.81.214419} {\bibfield  {journal} {\bibinfo  {journal} {Phys. Rev. B}\ }\textbf {\bibinfo {volume} {81}},\ \bibinfo {pages} {214419} (\bibinfo {year} {2010})}\BibitemShut {NoStop}%
\bibitem [{\citenamefont {Gao}\ \emph {et~al.}(2017)\citenamefont {Gao}, \citenamefont {Zaharko}, \citenamefont {Tsurkan}, \citenamefont {Su}, \citenamefont {White}, \citenamefont {Tucker}, \citenamefont {Roessli}, \citenamefont {Bourdarot}, \citenamefont {Sibille}, \citenamefont {Chernyshov}, \citenamefont {Fennell}, \citenamefont {Loidl},\ and\ \citenamefont {R{\"u}egg}}]{Gao17}%
  \BibitemOpen
  \bibfield  {author} {\bibinfo {author} {\bibfnamefont {S.}~\bibnamefont {Gao}}, \bibinfo {author} {\bibfnamefont {O.}~\bibnamefont {Zaharko}}, \bibinfo {author} {\bibfnamefont {V.}~\bibnamefont {Tsurkan}}, \bibinfo {author} {\bibfnamefont {Y.}~\bibnamefont {Su}}, \bibinfo {author} {\bibfnamefont {J.~S.}\ \bibnamefont {White}}, \bibinfo {author} {\bibfnamefont {G.}~\bibnamefont {Tucker}}, \bibinfo {author} {\bibfnamefont {B.}~\bibnamefont {Roessli}}, \bibinfo {author} {\bibfnamefont {F.}~\bibnamefont {Bourdarot}}, \bibinfo {author} {\bibfnamefont {R.}~\bibnamefont {Sibille}}, \bibinfo {author} {\bibfnamefont {D.}~\bibnamefont {Chernyshov}}, \bibinfo {author} {\bibfnamefont {T.}~\bibnamefont {Fennell}}, \bibinfo {author} {\bibfnamefont {A.}~\bibnamefont {Loidl}},\ and\ \bibinfo {author} {\bibfnamefont {C.}~\bibnamefont {R{\"u}egg}},\ }\bibfield  {title} {\bibinfo {title} {{Spiral spin-liquid and the emergence of a vortex-like state in MnSc$_2$S$_4$}},\ }\href {https://doi.org/10.1038/nphys3914} {\bibfield
  {journal} {\bibinfo  {journal} {Nature Physics}\ }\textbf {\bibinfo {volume} {13}},\ \bibinfo {pages} {157} (\bibinfo {year} {2017})}\BibitemShut {NoStop}%
\bibitem [{\citenamefont {Iqbal}\ \emph {et~al.}(2018)\citenamefont {Iqbal}, \citenamefont {M\"uller}, \citenamefont {Jeschke}, \citenamefont {Thomale},\ and\ \citenamefont {Reuther}}]{Iqbal18}%
  \BibitemOpen
  \bibfield  {author} {\bibinfo {author} {\bibfnamefont {Y.}~\bibnamefont {Iqbal}}, \bibinfo {author} {\bibfnamefont {T.}~\bibnamefont {M\"uller}}, \bibinfo {author} {\bibfnamefont {H.~O.}\ \bibnamefont {Jeschke}}, \bibinfo {author} {\bibfnamefont {R.}~\bibnamefont {Thomale}},\ and\ \bibinfo {author} {\bibfnamefont {J.}~\bibnamefont {Reuther}},\ }\bibfield  {title} {\bibinfo {title} {{Stability of the spiral spin liquid in ${\mathrm{MnSc}}_{2}{\mathrm{S}}_{4}$}},\ }\href {https://doi.org/10.1103/PhysRevB.98.064427} {\bibfield  {journal} {\bibinfo  {journal} {Phys. Rev. B}\ }\textbf {\bibinfo {volume} {98}},\ \bibinfo {pages} {064427} (\bibinfo {year} {2018})}\BibitemShut {NoStop}%
\bibitem [{\citenamefont {Buessen}\ \emph {et~al.}(2018)\citenamefont {Buessen}, \citenamefont {Hering}, \citenamefont {Reuther},\ and\ \citenamefont {Trebst}}]{Buessen18}%
  \BibitemOpen
  \bibfield  {author} {\bibinfo {author} {\bibfnamefont {F.~L.}\ \bibnamefont {Buessen}}, \bibinfo {author} {\bibfnamefont {M.}~\bibnamefont {Hering}}, \bibinfo {author} {\bibfnamefont {J.}~\bibnamefont {Reuther}},\ and\ \bibinfo {author} {\bibfnamefont {S.}~\bibnamefont {Trebst}},\ }\bibfield  {title} {\bibinfo {title} {{Quantum Spin Liquids in Frustrated Spin-1 Diamond Antiferromagnets}},\ }\href {https://doi.org/10.1103/PhysRevLett.120.057201} {\bibfield  {journal} {\bibinfo  {journal} {Phys. Rev. Lett.}\ }\textbf {\bibinfo {volume} {120}},\ \bibinfo {pages} {057201} (\bibinfo {year} {2018})}\BibitemShut {NoStop}%
\bibitem [{\citenamefont {Ghosh}\ \emph {et~al.}(2019)\citenamefont {Ghosh}, \citenamefont {Iqbal}, \citenamefont {M{\"u}ller}, \citenamefont {Ponnaganti}, \citenamefont {Thomale}, \citenamefont {Narayanan}, \citenamefont {Reuther}, \citenamefont {Gingras},\ and\ \citenamefont {Jeschke}}]{Ghosh19}%
  \BibitemOpen
  \bibfield  {author} {\bibinfo {author} {\bibfnamefont {P.}~\bibnamefont {Ghosh}}, \bibinfo {author} {\bibfnamefont {Y.}~\bibnamefont {Iqbal}}, \bibinfo {author} {\bibfnamefont {T.}~\bibnamefont {M{\"u}ller}}, \bibinfo {author} {\bibfnamefont {R.~T.}\ \bibnamefont {Ponnaganti}}, \bibinfo {author} {\bibfnamefont {R.}~\bibnamefont {Thomale}}, \bibinfo {author} {\bibfnamefont {R.}~\bibnamefont {Narayanan}}, \bibinfo {author} {\bibfnamefont {J.}~\bibnamefont {Reuther}}, \bibinfo {author} {\bibfnamefont {M.~J.~P.}\ \bibnamefont {Gingras}},\ and\ \bibinfo {author} {\bibfnamefont {H.~O.}\ \bibnamefont {Jeschke}},\ }\bibfield  {title} {\bibinfo {title} {Breathing chromium spinels: a showcase for a variety of pyrochlore heisenberg hamiltonians},\ }\href {https://doi.org/10.1038/s41535-019-0202-z} {\bibfield  {journal} {\bibinfo  {journal} {npj Quantum Materials}\ }\textbf {\bibinfo {volume} {4}},\ \bibinfo {pages} {63} (\bibinfo {year} {2019})}\BibitemShut {NoStop}%
\bibitem [{\citenamefont {Niggemann}\ \emph {et~al.}(2019)\citenamefont {Niggemann}, \citenamefont {Hering},\ and\ \citenamefont {Reuther}}]{Niggemann20}%
  \BibitemOpen
  \bibfield  {author} {\bibinfo {author} {\bibfnamefont {N.}~\bibnamefont {Niggemann}}, \bibinfo {author} {\bibfnamefont {M.}~\bibnamefont {Hering}},\ and\ \bibinfo {author} {\bibfnamefont {J.}~\bibnamefont {Reuther}},\ }\bibfield  {title} {\bibinfo {title} {{Classical spiral spin liquids as a possible route to quantum spin liquids}},\ }\href {https://doi.org/10.1088/1361-648X/ab4480} {\bibfield  {journal} {\bibinfo  {journal} {Journal of Physics: Condensed Matter}\ }\textbf {\bibinfo {volume} {32}},\ \bibinfo {pages} {024001} (\bibinfo {year} {2019})}\BibitemShut {NoStop}%
\bibitem [{\citenamefont {Yao}\ \emph {et~al.}(2021)\citenamefont {Yao}, \citenamefont {Liu}, \citenamefont {Huang}, \citenamefont {Wang},\ and\ \citenamefont {Chen}}]{Yao21}%
  \BibitemOpen
  \bibfield  {author} {\bibinfo {author} {\bibfnamefont {X.-P.}\ \bibnamefont {Yao}}, \bibinfo {author} {\bibfnamefont {J.~Q.}\ \bibnamefont {Liu}}, \bibinfo {author} {\bibfnamefont {C.-J.}\ \bibnamefont {Huang}}, \bibinfo {author} {\bibfnamefont {X.}~\bibnamefont {Wang}},\ and\ \bibinfo {author} {\bibfnamefont {G.}~\bibnamefont {Chen}},\ }\bibfield  {title} {\bibinfo {title} {{Generic spiral spin liquids}},\ }\href {https://doi.org/10.1007/s11467-021-1074-9} {\bibfield  {journal} {\bibinfo  {journal} {Frontiers of Physics}\ }\textbf {\bibinfo {volume} {16}},\ \bibinfo {pages} {53303} (\bibinfo {year} {2021})}\BibitemShut {NoStop}%
\bibitem [{\citenamefont {Gao}\ \emph {et~al.}(2022)\citenamefont {Gao}, \citenamefont {McGuire}, \citenamefont {Liu}, \citenamefont {Abernathy}, \citenamefont {Cruz}, \citenamefont {Frontzek}, \citenamefont {Stone},\ and\ \citenamefont {Christianson}}]{Gao22}%
  \BibitemOpen
  \bibfield  {author} {\bibinfo {author} {\bibfnamefont {S.}~\bibnamefont {Gao}}, \bibinfo {author} {\bibfnamefont {M.~A.}\ \bibnamefont {McGuire}}, \bibinfo {author} {\bibfnamefont {Y.}~\bibnamefont {Liu}}, \bibinfo {author} {\bibfnamefont {D.~L.}\ \bibnamefont {Abernathy}}, \bibinfo {author} {\bibfnamefont {C.~d.}\ \bibnamefont {Cruz}}, \bibinfo {author} {\bibfnamefont {M.}~\bibnamefont {Frontzek}}, \bibinfo {author} {\bibfnamefont {M.~B.}\ \bibnamefont {Stone}},\ and\ \bibinfo {author} {\bibfnamefont {A.~D.}\ \bibnamefont {Christianson}},\ }\bibfield  {title} {\bibinfo {title} {{Spiral Spin Liquid on a Honeycomb Lattice}},\ }\href {https://doi.org/10.1103/PhysRevLett.128.227201} {\bibfield  {journal} {\bibinfo  {journal} {Phys. Rev. Lett.}\ }\textbf {\bibinfo {volume} {128}},\ \bibinfo {pages} {227201} (\bibinfo {year} {2022})}\BibitemShut {NoStop}%
\bibitem [{\citenamefont {Graham}\ \emph {et~al.}(2023)\citenamefont {Graham}, \citenamefont {Qureshi}, \citenamefont {Ritter}, \citenamefont {Manuel}, \citenamefont {Wildes},\ and\ \citenamefont {Clark}}]{Graham23}%
  \BibitemOpen
  \bibfield  {author} {\bibinfo {author} {\bibfnamefont {J.~N.}\ \bibnamefont {Graham}}, \bibinfo {author} {\bibfnamefont {N.}~\bibnamefont {Qureshi}}, \bibinfo {author} {\bibfnamefont {C.}~\bibnamefont {Ritter}}, \bibinfo {author} {\bibfnamefont {P.}~\bibnamefont {Manuel}}, \bibinfo {author} {\bibfnamefont {A.~R.}\ \bibnamefont {Wildes}},\ and\ \bibinfo {author} {\bibfnamefont {L.}~\bibnamefont {Clark}},\ }\bibfield  {title} {\bibinfo {title} {{Experimental Evidence for the Spiral Spin Liquid in ${\mathrm{LiYbO}}_{2}$}},\ }\href {https://doi.org/10.1103/PhysRevLett.130.166703} {\bibfield  {journal} {\bibinfo  {journal} {Phys. Rev. Lett.}\ }\textbf {\bibinfo {volume} {130}},\ \bibinfo {pages} {166703} (\bibinfo {year} {2023})}\BibitemShut {NoStop}%
\bibitem [{\citenamefont {Yan}\ and\ \citenamefont {Reuther}(2022)}]{Yan22}%
  \BibitemOpen
  \bibfield  {author} {\bibinfo {author} {\bibfnamefont {H.}~\bibnamefont {Yan}}\ and\ \bibinfo {author} {\bibfnamefont {J.}~\bibnamefont {Reuther}},\ }\bibfield  {title} {\bibinfo {title} {{Low-energy structure of spiral spin liquids}},\ }\href {https://doi.org/10.1103/PhysRevResearch.4.023175} {\bibfield  {journal} {\bibinfo  {journal} {Phys. Rev. Res.}\ }\textbf {\bibinfo {volume} {4}},\ \bibinfo {pages} {023175} (\bibinfo {year} {2022})}\BibitemShut {NoStop}%
\bibitem [{\citenamefont {Pretko}(2017{\natexlab{a}})}]{Pretko17}%
  \BibitemOpen
  \bibfield  {author} {\bibinfo {author} {\bibfnamefont {M.}~\bibnamefont {Pretko}},\ }\bibfield  {title} {\bibinfo {title} {{Generalized electromagnetism of subdimensional particles: A spin liquid story}},\ }\href {https://doi.org/10.1103/PhysRevB.96.035119} {\bibfield  {journal} {\bibinfo  {journal} {Phys. Rev. B}\ }\textbf {\bibinfo {volume} {96}},\ \bibinfo {pages} {035119} (\bibinfo {year} {2017}{\natexlab{a}})}\BibitemShut {NoStop}%
\bibitem [{\citenamefont {Pretko}(2017{\natexlab{b}})}]{Pretko17b}%
  \BibitemOpen
  \bibfield  {author} {\bibinfo {author} {\bibfnamefont {M.}~\bibnamefont {Pretko}},\ }\bibfield  {title} {\bibinfo {title} {{Subdimensional particle structure of higher rank $U(1)$ spin liquids}},\ }\href {https://doi.org/10.1103/PhysRevB.95.115139} {\bibfield  {journal} {\bibinfo  {journal} {Phys. Rev. B}\ }\textbf {\bibinfo {volume} {95}},\ \bibinfo {pages} {115139} (\bibinfo {year} {2017}{\natexlab{b}})}\BibitemShut {NoStop}%
\bibitem [{\citenamefont {Prem}\ \emph {et~al.}(2018)\citenamefont {Prem}, \citenamefont {Vijay}, \citenamefont {Chou}, \citenamefont {Pretko},\ and\ \citenamefont {Nandkishore}}]{Prem18}%
  \BibitemOpen
  \bibfield  {author} {\bibinfo {author} {\bibfnamefont {A.}~\bibnamefont {Prem}}, \bibinfo {author} {\bibfnamefont {S.}~\bibnamefont {Vijay}}, \bibinfo {author} {\bibfnamefont {Y.-Z.}\ \bibnamefont {Chou}}, \bibinfo {author} {\bibfnamefont {M.}~\bibnamefont {Pretko}},\ and\ \bibinfo {author} {\bibfnamefont {R.~M.}\ \bibnamefont {Nandkishore}},\ }\bibfield  {title} {\bibinfo {title} {{Pinch point singularities of tensor spin liquids}},\ }\href {https://doi.org/10.1103/PhysRevB.98.165140} {\bibfield  {journal} {\bibinfo  {journal} {Phys. Rev. B}\ }\textbf {\bibinfo {volume} {98}},\ \bibinfo {pages} {165140} (\bibinfo {year} {2018})}\BibitemShut {NoStop}%
\bibitem [{\citenamefont {Hsieh}\ \emph {et~al.}(2022)\citenamefont {Hsieh}, \citenamefont {Ma},\ and\ \citenamefont {Radzihovsky}}]{PhysRevA.106.023321}%
  \BibitemOpen
  \bibfield  {author} {\bibinfo {author} {\bibfnamefont {T.-C.}\ \bibnamefont {Hsieh}}, \bibinfo {author} {\bibfnamefont {H.}~\bibnamefont {Ma}},\ and\ \bibinfo {author} {\bibfnamefont {L.}~\bibnamefont {Radzihovsky}},\ }\bibfield  {title} {\bibinfo {title} {{Helical superfluid in a frustrated honeycomb Bose-Hubbard model}},\ }\href {https://doi.org/10.1103/PhysRevA.106.023321} {\bibfield  {journal} {\bibinfo  {journal} {Phys. Rev. A}\ }\textbf {\bibinfo {volume} {106}},\ \bibinfo {pages} {023321} (\bibinfo {year} {2022})}\BibitemShut {NoStop}%
\bibitem [{\citenamefont {Zhai}\ and\ \citenamefont {Radzihovsky}(2021)}]{ZHAI2021168509}%
  \BibitemOpen
  \bibfield  {author} {\bibinfo {author} {\bibfnamefont {Z.}~\bibnamefont {Zhai}}\ and\ \bibinfo {author} {\bibfnamefont {L.}~\bibnamefont {Radzihovsky}},\ }\bibfield  {title} {\bibinfo {title} {{Fractonic gauge theory of smectics}},\ }\href {https://doi.org/https://doi.org/10.1016/j.aop.2021.168509} {\bibfield  {journal} {\bibinfo  {journal} {Annals of Physics}\ }\textbf {\bibinfo {volume} {435}},\ \bibinfo {pages} {168509} (\bibinfo {year} {2021})},\ \bibinfo {note} {special issue on Philip W. Anderson}\BibitemShut {NoStop}%
\bibitem [{\citenamefont {Radzihovsky}(2020)}]{PhysRevLett.125.267601}%
  \BibitemOpen
  \bibfield  {author} {\bibinfo {author} {\bibfnamefont {L.}~\bibnamefont {Radzihovsky}},\ }\bibfield  {title} {\bibinfo {title} {{Quantum Smectic Gauge Theory}},\ }\href {https://doi.org/10.1103/PhysRevLett.125.267601} {\bibfield  {journal} {\bibinfo  {journal} {Phys. Rev. Lett.}\ }\textbf {\bibinfo {volume} {125}},\ \bibinfo {pages} {267601} (\bibinfo {year} {2020})}\BibitemShut {NoStop}%
\bibitem [{\citenamefont {Radzihovsky}(2023)}]{radzihovsky2023critical}%
  \BibitemOpen
  \bibfield  {author} {\bibinfo {author} {\bibfnamefont {L.}~\bibnamefont {Radzihovsky}},\ }\href@noop {} {\bibinfo {title} {{Critical Matter}}} (\bibinfo {year} {2023}),\ \Eprint {https://arxiv.org/abs/2306.03142} {arXiv:2306.03142 [cond-mat.stat-mech]} \BibitemShut {NoStop}%
\bibitem [{\citenamefont {Machon}\ \emph {et~al.}(2019)\citenamefont {Machon}, \citenamefont {Aharoni}, \citenamefont {Hu},\ and\ \citenamefont {Kamien}}]{Machon2019}%
  \BibitemOpen
  \bibfield  {author} {\bibinfo {author} {\bibfnamefont {T.}~\bibnamefont {Machon}}, \bibinfo {author} {\bibfnamefont {H.}~\bibnamefont {Aharoni}}, \bibinfo {author} {\bibfnamefont {Y.}~\bibnamefont {Hu}},\ and\ \bibinfo {author} {\bibfnamefont {R.~D.}\ \bibnamefont {Kamien}},\ }\bibfield  {title} {\bibinfo {title} {{Aspects of Defect Topology in Smectic Liquid Crystals}},\ }\href {https://doi.org/10.1007/s00220-019-03366-y} {\bibfield  {journal} {\bibinfo  {journal} {Communications in Mathematical Physics}\ }\textbf {\bibinfo {volume} {372}},\ \bibinfo {pages} {525–542} (\bibinfo {year} {2019})}\BibitemShut {NoStop}%
\bibitem [{\citenamefont {Shimokawa}\ and\ \citenamefont {Kawamura}(2019)}]{Shimokawa19}%
  \BibitemOpen
  \bibfield  {author} {\bibinfo {author} {\bibfnamefont {T.}~\bibnamefont {Shimokawa}}\ and\ \bibinfo {author} {\bibfnamefont {H.}~\bibnamefont {Kawamura}},\ }\bibfield  {title} {\bibinfo {title} {{Ripple State in the Frustrated Honeycomb-Lattice Antiferromagnet}},\ }\href {https://doi.org/10.1103/PhysRevLett.123.057202} {\bibfield  {journal} {\bibinfo  {journal} {Phys. Rev. Lett.}\ }\textbf {\bibinfo {volume} {123}},\ \bibinfo {pages} {057202} (\bibinfo {year} {2019})}\BibitemShut {NoStop}%
\bibitem [{\citenamefont {Fisher}\ and\ \citenamefont {Selke}(1980)}]{Fisher80}%
  \BibitemOpen
  \bibfield  {author} {\bibinfo {author} {\bibfnamefont {M.~E.}\ \bibnamefont {Fisher}}\ and\ \bibinfo {author} {\bibfnamefont {W.}~\bibnamefont {Selke}},\ }\bibfield  {title} {\bibinfo {title} {{Infinitely Many Commensurate Phases in a Simple Ising Model}},\ }\href {https://doi.org/10.1103/PhysRevLett.44.1502} {\bibfield  {journal} {\bibinfo  {journal} {Phys. Rev. Lett.}\ }\textbf {\bibinfo {volume} {44}},\ \bibinfo {pages} {1502} (\bibinfo {year} {1980})}\BibitemShut {NoStop}%
\bibitem [{\citenamefont {Seifert}\ and\ \citenamefont {Vojta}(2019)}]{Seifert19}%
  \BibitemOpen
  \bibfield  {author} {\bibinfo {author} {\bibfnamefont {U.~F.~P.}\ \bibnamefont {Seifert}}\ and\ \bibinfo {author} {\bibfnamefont {M.}~\bibnamefont {Vojta}},\ }\bibfield  {title} {\bibinfo {title} {{Theory of partial quantum disorder in the stuffed honeycomb Heisenberg antiferromagnet}},\ }\href {https://doi.org/10.1103/PhysRevB.99.155156} {\bibfield  {journal} {\bibinfo  {journal} {Phys. Rev. B}\ }\textbf {\bibinfo {volume} {99}},\ \bibinfo {pages} {155156} (\bibinfo {year} {2019})}\BibitemShut {NoStop}%
\bibitem [{\citenamefont {Blesio}\ \emph {et~al.}(2023)\citenamefont {Blesio}, \citenamefont {Lisandrini},\ and\ \citenamefont {Gonzalez}}]{Blesio23}%
  \BibitemOpen
  \bibfield  {author} {\bibinfo {author} {\bibfnamefont {G.~G.}\ \bibnamefont {Blesio}}, \bibinfo {author} {\bibfnamefont {F.~T.}\ \bibnamefont {Lisandrini}},\ and\ \bibinfo {author} {\bibfnamefont {M.~G.}\ \bibnamefont {Gonzalez}},\ }\bibfield  {title} {\bibinfo {title} {{Partially disordered Heisenberg antiferromagnet with short-range stripe correlations}},\ }\href {https://doi.org/10.1103/PhysRevB.107.134418} {\bibfield  {journal} {\bibinfo  {journal} {Phys. Rev. B}\ }\textbf {\bibinfo {volume} {107}},\ \bibinfo {pages} {134418} (\bibinfo {year} {2023})}\BibitemShut {NoStop}%
\bibitem [{\citenamefont {Chandra}\ \emph {et~al.}(1990)\citenamefont {Chandra}, \citenamefont {Coleman},\ and\ \citenamefont {Larkin}}]{Chandra90}%
  \BibitemOpen
  \bibfield  {author} {\bibinfo {author} {\bibfnamefont {P.}~\bibnamefont {Chandra}}, \bibinfo {author} {\bibfnamefont {P.}~\bibnamefont {Coleman}},\ and\ \bibinfo {author} {\bibfnamefont {A.~I.}\ \bibnamefont {Larkin}},\ }\bibfield  {title} {\bibinfo {title} {{Ising transition in frustrated Heisenberg models}},\ }\href {https://doi.org/10.1103/PhysRevLett.64.88} {\bibfield  {journal} {\bibinfo  {journal} {Phys. Rev. Lett.}\ }\textbf {\bibinfo {volume} {64}},\ \bibinfo {pages} {88} (\bibinfo {year} {1990})}\BibitemShut {NoStop}%
\bibitem [{\citenamefont {Weber}\ \emph {et~al.}(2003)\citenamefont {Weber}, \citenamefont {Capriotti}, \citenamefont {Misguich}, \citenamefont {Becca}, \citenamefont {Elhajal},\ and\ \citenamefont {Mila}}]{Weber03}%
  \BibitemOpen
  \bibfield  {author} {\bibinfo {author} {\bibfnamefont {C.}~\bibnamefont {Weber}}, \bibinfo {author} {\bibfnamefont {L.}~\bibnamefont {Capriotti}}, \bibinfo {author} {\bibfnamefont {G.}~\bibnamefont {Misguich}}, \bibinfo {author} {\bibfnamefont {F.}~\bibnamefont {Becca}}, \bibinfo {author} {\bibfnamefont {M.}~\bibnamefont {Elhajal}},\ and\ \bibinfo {author} {\bibfnamefont {F.}~\bibnamefont {Mila}},\ }\bibfield  {title} {\bibinfo {title} {{Ising Transition Driven by Frustration in a 2D Classical Model with Continuous Symmetry}},\ }\href {https://doi.org/10.1103/PhysRevLett.91.177202} {\bibfield  {journal} {\bibinfo  {journal} {Phys. Rev. Lett.}\ }\textbf {\bibinfo {volume} {91}},\ \bibinfo {pages} {177202} (\bibinfo {year} {2003})}\BibitemShut {NoStop}%
\bibitem [{\citenamefont {Wu}(1982)}]{Wu82}%
  \BibitemOpen
  \bibfield  {author} {\bibinfo {author} {\bibfnamefont {F.~Y.}\ \bibnamefont {Wu}},\ }\bibfield  {title} {\bibinfo {title} {{The Potts model}},\ }\href {https://doi.org/10.1103/RevModPhys.54.235} {\bibfield  {journal} {\bibinfo  {journal} {Rev. Mod. Phys.}\ }\textbf {\bibinfo {volume} {54}},\ \bibinfo {pages} {235} (\bibinfo {year} {1982})}\BibitemShut {NoStop}%
\bibitem [{\citenamefont {Grison}\ \emph {et~al.}(2020)\citenamefont {Grison}, \citenamefont {Viot}, \citenamefont {Bernu},\ and\ \citenamefont {Messio}}]{Grison20}%
  \BibitemOpen
  \bibfield  {author} {\bibinfo {author} {\bibfnamefont {V.}~\bibnamefont {Grison}}, \bibinfo {author} {\bibfnamefont {P.}~\bibnamefont {Viot}}, \bibinfo {author} {\bibfnamefont {B.}~\bibnamefont {Bernu}},\ and\ \bibinfo {author} {\bibfnamefont {L.}~\bibnamefont {Messio}},\ }\bibfield  {title} {\bibinfo {title} {{Emergent Potts order in the kagome ${J}_{1}\ensuremath{-}{J}_{3}$ Heisenberg model}},\ }\href {https://doi.org/10.1103/PhysRevB.102.214424} {\bibfield  {journal} {\bibinfo  {journal} {Phys. Rev. B}\ }\textbf {\bibinfo {volume} {102}},\ \bibinfo {pages} {214424} (\bibinfo {year} {2020})}\BibitemShut {NoStop}%
\bibitem [{\citenamefont {Moessner}\ and\ \citenamefont {Chalker}(1998{\natexlab{b}})}]{Moessner98b}%
  \BibitemOpen
  \bibfield  {author} {\bibinfo {author} {\bibfnamefont {R.}~\bibnamefont {Moessner}}\ and\ \bibinfo {author} {\bibfnamefont {J.~T.}\ \bibnamefont {Chalker}},\ }\bibfield  {title} {\bibinfo {title} {{Low-temperature properties of classical geometrically frustrated antiferromagnets}},\ }\href {https://doi.org/10.1103/PhysRevB.58.12049} {\bibfield  {journal} {\bibinfo  {journal} {Phys. Rev. B}\ }\textbf {\bibinfo {volume} {58}},\ \bibinfo {pages} {12049} (\bibinfo {year} {1998}{\natexlab{b}})}\BibitemShut {NoStop}%
\bibitem [{\citenamefont {Conlon}\ and\ \citenamefont {Chalker}(2009)}]{Conlon09}%
  \BibitemOpen
  \bibfield  {author} {\bibinfo {author} {\bibfnamefont {P.~H.}\ \bibnamefont {Conlon}}\ and\ \bibinfo {author} {\bibfnamefont {J.~T.}\ \bibnamefont {Chalker}},\ }\bibfield  {title} {\bibinfo {title} {{Spin Dynamics in Pyrochlore Heisenberg Antiferromagnets}},\ }\href {https://doi.org/10.1103/PhysRevLett.102.237206} {\bibfield  {journal} {\bibinfo  {journal} {Phys. Rev. Lett.}\ }\textbf {\bibinfo {volume} {102}},\ \bibinfo {pages} {237206} (\bibinfo {year} {2009})}\BibitemShut {NoStop}%
\bibitem [{\citenamefont {Zhang}\ \emph {et~al.}(2019)\citenamefont {Zhang}, \citenamefont {Changlani}, \citenamefont {Plumb}, \citenamefont {Tchernyshyov},\ and\ \citenamefont {Moessner}}]{Zhang19}%
  \BibitemOpen
  \bibfield  {author} {\bibinfo {author} {\bibfnamefont {S.}~\bibnamefont {Zhang}}, \bibinfo {author} {\bibfnamefont {H.~J.}\ \bibnamefont {Changlani}}, \bibinfo {author} {\bibfnamefont {K.~W.}\ \bibnamefont {Plumb}}, \bibinfo {author} {\bibfnamefont {O.}~\bibnamefont {Tchernyshyov}},\ and\ \bibinfo {author} {\bibfnamefont {R.}~\bibnamefont {Moessner}},\ }\bibfield  {title} {\bibinfo {title} {{Dynamical Structure Factor of the Three-Dimensional Quantum Spin Liquid Candidate ${\mathrm{NaCaNi}}_{2}{\mathrm{F}}_{7}$}},\ }\href {https://doi.org/10.1103/PhysRevLett.122.167203} {\bibfield  {journal} {\bibinfo  {journal} {Phys. Rev. Lett.}\ }\textbf {\bibinfo {volume} {122}},\ \bibinfo {pages} {167203} (\bibinfo {year} {2019})}\BibitemShut {NoStop}%
\bibitem [{\citenamefont {Nho}\ and\ \citenamefont {Landau}(2002)}]{Nho02}%
  \BibitemOpen
  \bibfield  {author} {\bibinfo {author} {\bibfnamefont {K.}~\bibnamefont {Nho}}\ and\ \bibinfo {author} {\bibfnamefont {D.~P.}\ \bibnamefont {Landau}},\ }\bibfield  {title} {\bibinfo {title} {{Spin-dynamics simulations of the triangular antiferromagnetic XY model}},\ }\href {https://doi.org/10.1103/PhysRevB.66.174403} {\bibfield  {journal} {\bibinfo  {journal} {Phys. Rev. B}\ }\textbf {\bibinfo {volume} {66}},\ \bibinfo {pages} {174403} (\bibinfo {year} {2002})}\BibitemShut {NoStop}%
\bibitem [{\citenamefont {Toth}\ and\ \citenamefont {Lake}(2015)}]{Toth15}%
  \BibitemOpen
  \bibfield  {author} {\bibinfo {author} {\bibfnamefont {S.}~\bibnamefont {Toth}}\ and\ \bibinfo {author} {\bibfnamefont {B.}~\bibnamefont {Lake}},\ }\bibfield  {title} {\bibinfo {title} {{Linear spin wave theory for single-Q incommensurate magnetic structures}},\ }\href {https://doi.org/10.1088/0953-8984/27/16/166002} {\bibfield  {journal} {\bibinfo  {journal} {Journal of Physics: Condensed Matter}\ }\textbf {\bibinfo {volume} {27}},\ \bibinfo {pages} {166002} (\bibinfo {year} {2015})}\BibitemShut {NoStop}%
\bibitem [{\citenamefont {Kosterlitz}\ and\ \citenamefont {Thouless}(1973)}]{Kosterlitz73}%
  \BibitemOpen
  \bibfield  {author} {\bibinfo {author} {\bibfnamefont {J.~M.}\ \bibnamefont {Kosterlitz}}\ and\ \bibinfo {author} {\bibfnamefont {D.~J.}\ \bibnamefont {Thouless}},\ }\bibfield  {title} {\bibinfo {title} {{Ordering, metastability and phase transitions in two-dimensional systems}},\ }\href {https://doi.org/10.1088/0022-3719/6/7/010} {\bibfield  {journal} {\bibinfo  {journal} {Journal of Physics C: Solid State Physics}\ }\textbf {\bibinfo {volume} {6}},\ \bibinfo {pages} {1181} (\bibinfo {year} {1973})}\BibitemShut {NoStop}%
\bibitem [{\citenamefont {Kosterlitz}(1974)}]{Kosterlitz74}%
  \BibitemOpen
  \bibfield  {author} {\bibinfo {author} {\bibfnamefont {J.~M.}\ \bibnamefont {Kosterlitz}},\ }\bibfield  {title} {\bibinfo {title} {{The critical properties of the two-dimensional xy model}},\ }\href {https://doi.org/10.1088/0022-3719/7/6/005} {\bibfield  {journal} {\bibinfo  {journal} {Journal of Physics C: Solid State Physics}\ }\textbf {\bibinfo {volume} {7}},\ \bibinfo {pages} {1046} (\bibinfo {year} {1974})}\BibitemShut {NoStop}%
\bibitem [{\citenamefont {Ota}\ \emph {et~al.}(1992)\citenamefont {Ota}, \citenamefont {Ota},\ and\ \citenamefont {Fahnle}}]{Ota92}%
  \BibitemOpen
  \bibfield  {author} {\bibinfo {author} {\bibfnamefont {S.}~\bibnamefont {Ota}}, \bibinfo {author} {\bibfnamefont {S.~B.}\ \bibnamefont {Ota}},\ and\ \bibinfo {author} {\bibfnamefont {M.}~\bibnamefont {Fahnle}},\ }\bibfield  {title} {\bibinfo {title} {{Microcanonical Monte Carlo simulations for the two-dimensional XY model}},\ }\href {https://doi.org/10.1088/0953-8984/4/24/011} {\bibfield  {journal} {\bibinfo  {journal} {Journal of Physics: Condensed Matter}\ }\textbf {\bibinfo {volume} {4}},\ \bibinfo {pages} {5411} (\bibinfo {year} {1992})}\BibitemShut {NoStop}%
\bibitem [{\citenamefont {Lee}\ \emph {et~al.}(2005)\citenamefont {Lee}, \citenamefont {Lee},\ and\ \citenamefont {mook Kim}}]{Lee05}%
  \BibitemOpen
  \bibfield  {author} {\bibinfo {author} {\bibfnamefont {K.~W.}\ \bibnamefont {Lee}}, \bibinfo {author} {\bibfnamefont {C.~E.}\ \bibnamefont {Lee}},\ and\ \bibinfo {author} {\bibfnamefont {I.}~\bibnamefont {mook Kim}},\ }\bibfield  {title} {\bibinfo {title} {{Helicity modulus and vortex density in the two-dimensional easy-plane Heisenberg model on a square lattice}},\ }\href {https://doi.org/https://doi.org/10.1016/j.ssc.2005.03.054} {\bibfield  {journal} {\bibinfo  {journal} {Solid State Communications}\ }\textbf {\bibinfo {volume} {135}},\ \bibinfo {pages} {95} (\bibinfo {year} {2005})}\BibitemShut {NoStop}%
\bibitem [{\citenamefont {Nguyen}\ and\ \citenamefont {Boninsegni}(2021)}]{Nguyen21}%
  \BibitemOpen
  \bibfield  {author} {\bibinfo {author} {\bibfnamefont {P.~H.}\ \bibnamefont {Nguyen}}\ and\ \bibinfo {author} {\bibfnamefont {M.}~\bibnamefont {Boninsegni}},\ }\bibfield  {title} {\bibinfo {title} {{Superfluid Transition and Specific Heat of the 2D x-y Model: Monte Carlo Simulation}},\ }\href {https://doi.org/https://doi.org/10.3390/app11114931} {\bibfield  {journal} {\bibinfo  {journal} {Applied Sciences}\ }\textbf {\bibinfo {volume} {11}},\ \bibinfo {pages} {4931} (\bibinfo {year} {2021})}\BibitemShut {NoStop}%
\bibitem [{\citenamefont {Ma}\ and\ \citenamefont {Pretko}(2018)}]{PhysRevB.98.125105}%
  \BibitemOpen
  \bibfield  {author} {\bibinfo {author} {\bibfnamefont {H.}~\bibnamefont {Ma}}\ and\ \bibinfo {author} {\bibfnamefont {M.}~\bibnamefont {Pretko}},\ }\bibfield  {title} {\bibinfo {title} {Higher-rank deconfined quantum criticality at the lifshitz transition and the exciton bose condensate},\ }\href {https://doi.org/10.1103/PhysRevB.98.125105} {\bibfield  {journal} {\bibinfo  {journal} {Phys. Rev. B}\ }\textbf {\bibinfo {volume} {98}},\ \bibinfo {pages} {125105} (\bibinfo {year} {2018})}\BibitemShut {NoStop}%
\bibitem [{\citenamefont {Ferdinand}\ and\ \citenamefont {Fisher}(1969)}]{Ferdinand69}%
  \BibitemOpen
  \bibfield  {author} {\bibinfo {author} {\bibfnamefont {A.~E.}\ \bibnamefont {Ferdinand}}\ and\ \bibinfo {author} {\bibfnamefont {M.~E.}\ \bibnamefont {Fisher}},\ }\bibfield  {title} {\bibinfo {title} {{Bounded and Inhomogeneous Ising Models. I. Specific-Heat Anomaly of a Finite Lattice}},\ }\href {https://doi.org/10.1103/PhysRev.185.832} {\bibfield  {journal} {\bibinfo  {journal} {Phys. Rev.}\ }\textbf {\bibinfo {volume} {185}},\ \bibinfo {pages} {832} (\bibinfo {year} {1969})}\BibitemShut {NoStop}%
\bibitem [{\citenamefont {Landau}(1976)}]{Landa76}%
  \BibitemOpen
  \bibfield  {author} {\bibinfo {author} {\bibfnamefont {D.~P.}\ \bibnamefont {Landau}},\ }\bibfield  {title} {\bibinfo {title} {{Finite-size behavior of the Ising square lattice}},\ }\href {https://doi.org/10.1103/PhysRevB.13.2997} {\bibfield  {journal} {\bibinfo  {journal} {Phys. Rev. B}\ }\textbf {\bibinfo {volume} {13}},\ \bibinfo {pages} {2997} (\bibinfo {year} {1976})}\BibitemShut {NoStop}%
\bibitem [{\citenamefont {Privman}(1990)}]{Privman90}%
  \BibitemOpen
  \bibfield  {author} {\bibinfo {author} {\bibfnamefont {V.}~\bibnamefont {Privman}},\ }\href {https://doi.org/10.1142/1011} {\emph {\bibinfo {title} {{Finite Size Scaling and Numerical Simulation of Statistical Systems}}}}\ (\bibinfo  {publisher} {WORLD SCIENTIFIC},\ \bibinfo {year} {1990})\BibitemShut {NoStop}%
\bibitem [{\citenamefont {Savitzky}\ and\ \citenamefont {Golay}(1964)}]{Savitzky64}%
  \BibitemOpen
  \bibfield  {author} {\bibinfo {author} {\bibfnamefont {A.}~\bibnamefont {Savitzky}}\ and\ \bibinfo {author} {\bibfnamefont {M.~J.~E.}\ \bibnamefont {Golay}},\ }\bibfield  {title} {\bibinfo {title} {{Smoothing and Differentiation of Data by Simplified Least Squares Procedures.}},\ }\href {https://doi.org/10.1021/ac60214a047} {\bibfield  {journal} {\bibinfo  {journal} {Analytical Chemistry}\ }\textbf {\bibinfo {volume} {36}},\ \bibinfo {pages} {1627} (\bibinfo {year} {1964})}\BibitemShut {NoStop}%
\end{thebibliography}%

\end{document}